\documentclass[preprint]{aastex}

\newcommand {\kms}{km~s$^{-1}$}
\newcommand {\xper}{$\chi$~Per}
\newcommand {\hper}{h~Per}
\newcommand {\hi}{H\,{\sc i}}
\newcommand {\sod}{Na\,{\sc i}}
\newcommand {\sodd}{Na\,{\sc i}~D}
\newcommand {\pot}{K\,{\sc i}}
\newcommand {\sodi}{Na\,{\sc i}~D$_1$}
\newcommand {\sodii}{Na\,{\sc i}~D$_2$}
\newcommand {\caii}{Ca\,{\sc ii}}
\newcommand {\caiik}{Ca\,{\sc ii}~K}
\newcommand {\caiih}{Ca\,{\sc ii}~H}
\newcommand {\hii}{H\,{\sc ii}}

\newcommand {\fsec}{\mbox{$. \! ^{\rm s}$}}
\newcommand {\fdeg}{\mbox{$. \! ^{\circ}$}}
\def\farcm  {\hbox{$.\mkern-4mu^\prime$}}
\newbox\grsign \setbox\grsign=\hbox{$>$} \newdimen\grdimen \grdimen=\ht\grsign
\newbox\simlessbox \newbox\simgreatbox
\setbox\simgreatbox=\hbox{\raise.5ex\hbox{$>$}\llap
     {\lower.5ex\hbox{$\sim$}}}\ht1=\grdimen\dp1=0pt
\setbox\simlessbox=\hbox{\raise.5ex\hbox{$<$}\llap
     {\lower.5ex\hbox{$\sim$}}}\ht2=\grdimen\dp2=0pt
\def\simgreat{\mathrel{\copy\simgreatbox}}
\def\simless{\mathrel{\copy\simlessbox}}

\begin{document}


\title{The Complex Interstellar Na\,{\sc i} Absorption toward h and $\chi$
Persei}


\author{Sean D. Points$^{1,2}$, James T. Lauroesch$^1$, and
David M. Meyer$^{1}$}

\affil{Northwestern University, Department of Physics and Astronomy, 
2145 Sheridan Road, Evanston, IL 60208} 

\email{spoints@ctio.noao.edu, jtl@elvis.astro.northwestern.edu,
davemeyer@northwestern.edu}


\altaffiltext{1}{Visiting Astronomer, Kitt Peak National Observatory (KPNO).
KPNO is a part of the National Optical Astronomy Observatories (NOAO), which
is operated by the Association of Universities for Research in Astronomy, 
Inc. (AURA) under contract with the National Science Foundation (NSF).}

\altaffiltext{2}{Current Address: Cerro Tololo Inter-American Observatory, 
Casilla 603, La Serena, Chile}

\begin{abstract}

Recent high spatial and spectral resolution investigations of the
diffuse interstellar medium (ISM) have found significant evidence for
small-scale variations in the interstellar gas on scales $\leq 1$~pc.
To better understand the nature of small-scale variations in the ISM,
we have used the KPNO WIYN Hydra multi-object spectrograph, which has
a mapping advantage over the single-axis, single-scale limitations of
studies using high proper motion stars and binary stars, to obtain
moderate resolution ($\sim 12$~\kms) interstellar \sodd\ absorption
spectra of 172 stars toward the double open cluster h and
$\chi$~Persei.  All of the sightlines toward the 150 stars with
spectra that reveal absorption from the Perseus spiral arm show
different interstellar \sodd\ absorption profiles in the Perseus arm
gas.  Additionally, we have utilized the KPNO Coud\'e Feed
spectrograph to obtain high-resolution ($\sim 3$~\kms) interstellar
\sodd\ absorption spectra of 24 of the brighter stars toward h and
\xper.  These spectra reveal an even greater complexity in the
interstellar \sodd\ absorption in the Perseus arm gas and show
individual components changing in number, velocity, and strength from
sightline to sightline.  If each of these individual velocity
components represents an isolated cloud, then it would appear that the
ISM of the Perseus arm gas consists of many small clouds.  Although
the absorption profiles vary even on the smallest scales probed by
these high-resolution data ($\sim 30''$; $\sim 0.35$~pc), our analysis
reveals that some interstellar \sodd\ absorption components from
sightline to sightline are related, implying that the ISM toward h and
\xper\ is probably comprised of sheets of gas in which we detect
variations due to differences in the local physical conditions of the
gas.

\end{abstract}



\section{Introduction}

Observations of optical interstellar absorption lines toward
early-type Galactic stars have long been used to investigate the
structure of the interstellar medium (ISM) in the Milky Way.  For
example, M\"unch (1953; 1957) examined interstellar \sod\ and \caii\
absorption lines in the spectra of 112 distant, bright, early-type
stars to determine if the velocity profiles showed any systematic
features that would provide evidence for the large-scale distribution
of interstellar clouds.  These observations revealed substantial
concentrations of interstellar gas toward Perseus and Cepheus that
were coincident with the recently recognized spiral arm features of
the Milky Way (Morgan, Sharpless, \& Osterbrock 1952).  In addition 
to recognizing the advantage of using interstellar absorption lines
toward clusters and associations to map large-scale ($\simgreat
1000$~pc) interstellar structures in the Galaxy, M\"unch (1953) also
remarked that interstellar absorption line spectra show appreciable
variation for stars within the same cluster.  Thus, interstellar
absorption lines could also be used to examine the small-scale 
variations of interstellar structures ($\simless 1$~pc) in the diffuse 
ISM.

Recently, high spatial resolution radio and optical studies of the
diffuse ISM have found significant evidence for small-scale structures
in the ISM.  
Observations of \hi\ 21-cm absorption toward extended extragalactic
radio sources (Dieter, Welch, \& Romney 1976; Diamond et al. 1989;
Faison et al. 1998; Faison \& Goss 2001) and high-velocity pulsars
(Frail et al. 1994) have suggested the presence of structure in
diffuse clouds on scales of tens of AUs.  The ubiquity of such
structures, however, has been called into question (Deshpande 2000;
Johnston et al. 2003; Stanimirovi\'c et al. 2003).  High-resolution
21-cm emission-line aperture synthesis maps reveal a diverse
population of \hi\ self-absorbing clouds with apparent diameters $\le
0.6$~pc in the Perseus spiral arm (Gibson et al. 2000).
In the optical, studies have focused on high proper motion stars
(Crawford et al. 2000; Lauroesch, Meyer, \& Blades 2000; Welty \&
Fitzpatrick 2001; Crawford 2002; Lauroesch \& Meyer 2003; Rollinde et
al. 2003), binary stars (Meyer \& Blades 1996; Watson \& Meyer 1996;
Lauroesch et al.  1998; Lauroesch \& Meyer 1999; Price, Crawford, \&
Barlow 2000; Price, Crawford, \& Howarth 2001a), and globular clusters
(Bates, Kemp, \& Montgomery 1993; Kennedy, Bates et al. 1995; Kennedy,
Bates, \& Kemp 1996; Kennedy, Bates, \& Kemp 1998; Meyer \& Lauroesch
1999; Andrews, Meyer, \& Lauroesch 2001; Smoker et al. 2001),
typically using the \sodd, \pot, and \caiik\ lines.  These
observations have revealed interstellar structure on scales of
$\approx 10$~AU for high proper motion stars, $\approx 10^2 - 10^4$~AU
for binary stars, and $\approx 10^3 - 10^6$~AU for globular clusters.

%{\it Theory Section}

At present, the precise nature of these small-scale features is 
not well understood.  Assuming spherical geometries, the \hi\
volume densities of the small-scale clouds detected in 21-cm 
absorption are $n$(\hi) $\simgreat 10^4$~cm$^{-3}$ with inferred 
thermal pressures of P$_{\rm Th} \sim 10^6$~cm$^{-3}$~K (Heiles 
1997).  Thus, these small-scale features are significantly 
over-pressured with respect to the ambient ISM and should dissipate on 
time-scales of $\sim 100$~yrs.  Consequently, they should not be 
a frequently detected component of the ISM.  In order to resolve
this discrepancy, several alternatives have been proposed.  
Heiles (1997) suggested that the small-scale absorbing features
may be cylinders seen face-on or disks seen edge-on, instead of
having spherical geometries.  Deshpande (2000) has suggested that
the small-scale \hi\ absorbers represent the tail-end of hierarchical
structure in the ISM and that the measured transverse size of the
absorbing clouds is not necessarily indicative of the depth through 
the absorbing clouds. 
In the optical, studies of small-scale structure in the ISM using 
interstellar \sodd\ absorption are hindered by the fact that \sod\ 
is not a dominant ion in diffuse clouds.  Thus, it is possible that 
the structures detected using interstellar \sodd\ absorption are not 
changes in the physical structure of the gas but rather caused by 
fluctuations in the physical conditions of the gas such as cloud 
temperature, pressure, electron density, and/or radiation field. 
Thus, it is uncertain if the variations detected on small scales
are caused by changes in the density of the interstellar gas, changes
in the ionization balance, cloud geometry or a combination of all 
these effects (Pan, Federman, \& Welty 2001).

In order to further investigate the small-scale structure of the ISM,
we have used the Hydra multi-object spectrograph on the 3.5-m
WIYN\footnote{The WIYN Observatory is a joint facility of the
University of Wisconsin-Madison, Indiana University, Yale University,
and the National Optical Astronomy Observatories.}  at Kitt Peak
National Observatory (KPNO) to obtain moderate resolution ($\sim
12$~\kms) interstellar \sodd\ absorption spectra of 172 stars toward
the double open cluster h and $\chi$~Persei (NGC~869 and NGC 884,
respectively).  This sightline is interesting because it passes
through higher velocity gas ($-75 \le V_{Hel} \le -20$~\kms) associated
with the Perseus spiral arm (M\"unch 1957) as well as the more local
gas associated with the Orion spiral arm.  Furthermore, we have
obtained high-resolution ($\sim 3$~\kms) interstellar \sodd\
absorption spectra toward 24 stars in the double open cluster.  Using
stars in an open cluster or OB association as background sources to
probe intervening small-scale structure has a mapping advantage over
the single-axis, single-scale limitation of a high-proper motion or
binary stars.  In effect, it is possible to obtain an absorption line
map of interstellar \sod\ using this method of multi-object spectroscopy.

This paper is organized as follows.  Section 2 provides details of the
observations that are used for this study and their reductions.  The
analysis of these data is described in \S 3.  We interpret our results 
in \S 4.  In \S 5, we summarize our results.

\section{Observations and Data Reduction}

\subsection{KPNO WIYN Hydra Multi-Object Spectra}

The KPNO WIYN observations of h and $\chi$ Per were obtained in 1998
September and October using the Hydra multi-object spectrograph in
queue mode.  Using the red cable fibers (2$\arcsec$ diameter), Hydra
can observe up to about 90 objects at a time over a 60$\arcmin$
diameter field with a minimum fiber separation of 37$\arcsec$.  Two
Hydra configurations per cluster were observed in order to circumvent
this minimum spacing limitation.  As shown in Figure~1, a total of 172
stars were observed with angular separations ranging from
11.5$\arcsec$ to 4950$\arcsec$ which would correspond to linear
separations ranging from about 0.1 pc to 50 pc at the double cluster
distance of $\approx$2000~pc.  The spectrograph was configured with
the Red Bench camera, a Tek 2048 CCD (T2KC), the echelle grating, and
an interference filter (X17) providing spectral coverage from 5725 to
5965 \AA\ at a 2.0 pixel resolution of 0.24 \AA\ or 12~\kms.  We
present the 150 moderate resolution (12~\kms) spectra along the
sightlines that show interstellar \sodii\ absorption in Perseus spiral
arm gas in Figure~2.

Utilizing this instrumental setup, net exposures ranging from 1800 to
3600 s were taken of the h and $\chi$ Per stars under clear sky
conditions characterized by 1--2$\arcsec$ seeing.  The individual CCD
frames were background-corrected, flat-fielded, combined, and
wavelength-calibrated using the NOAO IRAF\footnote{IRAF is distributed
by the National Optical Astronomy Observatories, which are operated by
the Association of Universities for Research in Astronomy, Inc., under
cooperative agreement with the National Science Foundation.} data
reduction package in order to extract the net stellar spectrum yielded
by each illuminated fiber.  Based on previous observations with this
spectrograph and on data comparisons with the KPNO Coud\'e Feed
spectrograph, the uncertainty in the zero level of these spectra due
to uncorrected scattered-light effects should be less than 3\%.  In
order to remove the telluric absorption in the vicinity of the \sodii\
$\lambda$5889.951 and D$_1$ $\lambda$5895.924 lines, these spectra
were all divided by an atmospheric template based on observations of
several rapidly rotating early-type stars with little intervening
interstellar matter.  The resulting Na I spectra of the 172 stars
observed toward h and $\chi$ Per have pixel-to-pixel S/N ratios
ranging from $\approx$40 to $\approx$300.

\subsection{KPNO Coud\'e Feed Spectra}

The KPNO Coud\'{e} Feed spectrograph observations of stars in h and
$\chi$ Per were obtained in 2001 November and 2002 November using
Camera 5 with the echelle grating and a Ford 3000$\times$1000 pixel
CCD chip. The resolution of the data was measured using the ThAr lamp
emission lines and is $\sim$3.4 km/s at the position of the \sodd\
lines.  A total of 24 stars were observed with spectra covering the
region between $\sim$5400 and 6800\AA.  Except for the very brightest
stars (for which exposure times of 1800 and 2700s were used)
individual exposures were 3600s long, with total exposure times
ranging up to 5 hours.  For the majority of the stars exposures were
taken on different nights at different grating tilts to reduce the
effect of any flaws in the CCD chip.  However, due to poor weather it
was not possible to get more than one exposure for 6 of the brighter
stars in the sample.  Exposures of blank sky were obtained during each
run for background subtraction purposes, and in addition observations
of the stars $\alpha$ Aql and $\alpha$ Leo were obtained each night as
a template for dividing out telluric absorption.  These 24 stars were
selected from the larger WIYN sample to provide good coverage of the
clusters, and because they had sufficiently large ultra-violet fluxes
(as well as large $v \cdot sin\ i$) to be observed using the
Far-Ultraviolet Spectroscopic Explorer ({\it FUSE}) satellite.

Reduction of the data was done with the NOAO IRAF echelle data
reduction package. The individual frames were first bias-subtracted
and flat-fielded, then the scattered light was removed.  The
individual orders were then optimally extracted and cleaned of
cosmic-ray hits using the variance weighted extraction method (Horne
1986; Marsh 1989).  All of the resulting one-dimensional spectra then
had the night sky background (including \sodd\ emission lines) removed
using appropriately scaled exposures of blank sky taken during each
run.  The sky-subtracted one-dimensional spectra were then wavelength
calibrated, corrected for telluric absorption, shifted to heliocentric
coordinates, summed, and finally continuum fitted using low-order
polynomials. Figures~3 and 4 show the final \sodii\ line profiles.
The resulting signal-to-noise (S/N) ratios for these observations are
typically 40--50 per pixel, with two faint stars having S/N-ratios of
30--35, and some of the brighter stars having S/N-ratios of 60--80 per
pixel.

\section{Analysis}

Our analysis includes a discussion of the identification of stars used
as background sources in our sample and an assessment of their cluster
membership.  We then describe the observed trends in the interstellar
\sodd\ absorption of the Perseus spiral arm gas ($-75 \le V_{Hel} \le
-20$~\kms) toward h and $\chi$~Per as revealed by our moderate
resolution WIYN data and our high-resolution Coud\'e Feed data.  In
the discussion and figures we primarily focus on examining the
interstellar Na I D2 absorption toward h and X Per in order to better
show the weaker features in the gas associated with the Perseus spiral
arm, although in our analysis we have also considered the weaker D1
profile.

\subsection{Stellar Identification}

Before identifying the stars in our WIYN sample, we first examined the
spectra and removed the stars which showed no interstellar \sodd\
absorption in the Perseus arm gas.  After this was completed, 150 of
172 stars remained in our sample.  The identification of these 150
stars was then made by searching the SIMBAD database for stars within
a 1$''$ radius of the positions of the $2''$ Hydra spectrograph fibers
and yielded the identities of 88 stars in our sample.  This search was
subsequently repeated using search radii of 2$''$ and 4$''$.  Even
when the search radius relaxed to 4$''$, the majority of stars (125
stars) only have one designation in the SIMBAD database.  The
remaining stars have either two (22 stars) or three (3 stars) possible
designations.  In order to determine which of the multiple stellar
identifications were more likely, we examined the Digital Sky Survey
plates toward h and $\chi$~Per and determined the stellar
identification based on the positions and brightness of the stars.

After the stars used as background sources in this study were
identified, we investigated the cluster membership of the stars by
comparing our list of identifications with that of Slesnick,
Hillenbrand, \& Massey (2002).  In this manner, we determined a
cluster membership of h~Per, $\chi$~Per, or neither for the stars in
our sample.  For stars in our sample that Slesnick et al. (2002) do
not recognize as members of either h~Per or $\chi$~Per, we assigned
the stars to either an h~Per or $\chi$~Per ``field'' based on the
stars' projected distances to the cluster cores (\hper: $\alpha_{\rm
J2000} = 2^{\rm h} 19^{\rm m} 22\fsec2$, $\delta_{\rm J2000} =
+57\arcdeg 09\arcmin 00\arcsec$; \xper: $\alpha_{\rm J2000} = 2^{\rm
h} 22^{\rm m} 12\fsec0$, $\delta_{\rm J2000} = +57\arcdeg 07\arcmin
12\arcsec$).  This resulted in 74 stars in our sample as being
projected toward h~Per and 76 stars lying projected toward $\chi$~Per.

\subsection{Moderate Resolution Spectra}

After stars in our moderate ($\sim 12$~\kms) resolution Hydra sample
were identified and assigned to one of the clusters, we can begin to
examine the interstellar \sodii\ (5889\AA) absorption toward each
cluster.  To investigate the variations in interstellar \sodii\
absorption toward each cluster, we sorted the spectra into declination
bins and arranged them from east to west, roughly corresponding to
their position on the sky.  We present the interstellar \sodii\
absorption spectra toward $\chi$ and h~Per in Figures~2a \& b,
respectively.  In Figure~2 we have labeled the spectra with the
appropriate stellar identifications from the Henry Draper (HD; Cannon
\& Pickering 1918), Bonner Durchmusterung (BD; Argelander 1903), and
Oosterhoff (Oo; Oosterhoff 1937) catalogs.  The eleven unlabeled
spectra in Figure~2 have identifications, but they are not listed in
the three aforementioned catalogs.  Where applicable, we have
designated the cluster memberships of stars in Figure~2, as determined
by Slesnick et al. (2002), and also labeled stars within a $5'$ radius
of the respective cluster cores.

Our Hydra interstellar \sodii\ absorption line profiles reveal a very
complex velocity structure from point-to-point over the spatial extent
of $\chi$ and \hper.  In addition to the saturated interstellar
\sodii\ absorption by local, low velocity gas ($-20 \le V_{Hel} \le 
+10$~\kms) in the Orion spiral arm, the spectra toward $\chi$ and \hper\ 
show interstellar \sodii\ absorption at intermediate ($-55 \le V_{Hel} \le
-20$~\kms) and higher ($-70 \le V_{Hel} \le -55$~\kms) velocities that
is associated with the Perseus spiral arm.  As seen in Figure~2, all of 
these spectra show different interstellar \sodii\ absorption profiles for 
gas in the Perseus spiral arm, i.e., no two profiles are identical.  

The moderate resolution interstellar \sodii\ absorption line spectra
toward $\chi$ and \hper\ can be classified as having one of three
different shapes: (1) blended with no distinct peaks in the Perseus
arm gas (e.g., Oo\,2794, Oo\,892); (2) one distinct peak in the
Perseus arm gas with an extended wing suggesting more components
(e.g., BD$+$56\arcdeg 560, HD\,14162); and (3) two distinct peaks in
the Perseus arm gas (e.g., BD$+$56\arcdeg 610, HD\,14186).  The vast
majority of the absorption line spectra ($\sim 87\%$) toward h and
\xper\ fall into the second classification category.  As we discuss in
\S3.3, higher resolution ($\sim 3$~\kms) KPNO Coud\'e Feed spectra 
reveal a multitude of interstellar \sodii\ absorption components 
along any particular sightline.  Therefore, some care must be exercised 
to avoid over-interpreting the moderate resolution ($\sim 12$~\kms) 
spectra.  We discuss the moderate resolution interstellar \sodii\ 
absorption spectra toward each cluster in more detail below.

\subsubsection{$\chi$~Per}

The interstellar \sodii\ absorption line profiles toward \xper\ 
(Figure~2a) reveal a very chaotic structure in terms of both velocity 
and intensity variations in the Perseus arm gas.  As previously 
mentioned, we have arranged the spectra in Figure~2a so that their 
location in the figure approximates the spatial distribution of the 
stars toward the cluster.  Thus, we will discuss variations in the
interstellar \sodii\ absorption line profiles in terms of their strength, 
or optical depth ($\tau_{V} = -ln[I_{V}]$), in a given velocity 
range as a function of their spatial distribution across the cluster.

In the northeastern region of \xper, the interstellar \sodii\
absorption profiles are asymmetric.  The interstellar absorption
associated with the higher velocity ($-70 \le V_{Hel} \le -55$~\kms) 
gas appears as an extended blue wing in the line profile and the 
intermediate velocity ($-55 \le V_{Hel} \le -20$~\kms) gas appears 
as a bump on the line profile between the higher velocity component 
and the saturated low velocity Orion arm gas.  In contrast, the 
interstellar \sodii\ absorption of Perseus arm gas is more apparent 
toward the rest of the stars in this field.  Toward the core of 
\xper\ the interstellar \sodii\ absorption is stronger in both the 
higher and intermediate velocity components of the Perseus arm gas.  
Consequently, the majority of spectra along these sightlines seem 
to be double-peaked where the higher and intermediate velocity Perseus 
arm components appear blended together as one component and the low 
velocity Orion arm gas appears as the other component.  An examination 
of Figure~2a reveals one trend in these ``double-peaked'' spectra.  
In the upper left quadrant of Figure~2a (northeast region of \xper), 
the blended Perseus arm component has a local minimum to the blue of 
$-50$~\kms.  In the lower right section of Figure~2a (southwest region 
of \xper), the local minimum lies to the red of $-50$~\kms.  This 
suggests that absorption due to the intermediate velocity gas in the 
Perseus arm becomes stronger with respect to the higher velocity Perseus 
arm gas in the southwest.  The interstellar \sodii\ absorption profiles 
toward \xper\ that show two distinct peaks in the Perseus arm gas in 
addition to the low velocity Orion arm gas (e.g., BD$+$56\arcdeg 549 and 
BD$+$56\arcdeg 610) are primarily located away from the cluster core.

\subsubsection{h~Per}

Our moderate resolution interstellar \sodii\ absorption line 
spectra toward \hper\ (Figure~2b) also reveal complex structure 
in the Perseus arm gas.  Similar to our analysis of the interstellar 
\sodii\ absorption line profiles toward \xper, we have arranged the 
spectra shown in Figure~2b so that their location in the figure 
approximates the spatial distribution of our sightlines toward 
the cluster.

Although the interstellar \sodii\ absorption profiles toward 
\hper\ (Figure~2b) are similar to those seen toward \xper\ 
(Figure~2a), they generally show stronger interstellar \sodii\ 
absorption in the Perseus arm gas.  In the northern and eastern 
regions of \hper\ the line profiles reveal very strong interstellar 
\sodii\ absorption where the higher and intermediate velocity Perseus 
arm gas are blended into one component that has a line depth 
comparable to the Orion arm gas.  Furthermore, there is little 
separation between the peak of the blended Perseus arm component 
and the Orion arm gas, indicating a substantial amount of intermediate
velocity Perseus arm gas.  Toward the core of \hper\ a noticeable 
separation appears between the Perseus arm gas and the Orion arm 
gas and the peak of the higher and intermediate velocity gas blend 
has shifted to the blue, suggesting that the intermediate velocity 
Perseus arm component is decreasing in strength with respect to the 
higher velocity Perseus arm gas.  This apparent trend continues along 
the southwestern periphery of \hper. The gap between the Perseus arm 
gas and the Orion arm gas is even more pronounced than toward the core 
of \hper.  It is also apparent in Figure~2b that the interstellar 
\sodii\ absorption in the higher velocity Perseus arm gas in the 
southwestern region is strong and is very likely saturated. 

\subsection{High Resolution Spectra}

In order to better understand the variations in intensity and velocity
of interstellar \sodii\ absorption, we have also obtained
high-resolution ($\sim 3$~\kms) spectra toward 24 of the brightest
stars projected along the line-of-sight to h and \xper.  In Figure~3,
we present the high-resolution interstellar \sodii\ absorption spectra
of stars toward h and \xper.  We have arranged the spectra in Figure~3
to show the relative positions of the background target stars, similar
to our method of analysis for the moderate resolution data.  The
coordinate offsets provided in the lower left of each panel indicate
the relative position of any particular sightline with respect to
HD\,14250, which lies approximately between the cluster cores.  Thus,
stars to the west of HD\,14250 are projected toward \hper\ and stars
to the east of it are projected toward \xper.  To aid in the
comparison between the high-resolution and moderate resolution data,
we have plotted the high-resolution spectra in Figure~3 on the same
intensity and velocity scale as the spectra presented in Figure~2.

The interstellar \sodii\ absorption line spectra presented in Figure~3
show the vast complexity of interstellar gas toward h and \xper\ that
was first noted by M\"unch (1953; 1957).  These high-resolution data
reveal variations in interstellar \sodii\ absorption in an even more
dramatic fashion than the moderate resolution data.  The individual
components change in number, velocity, and strength from sightline to
sightline.  For example, the moderate resolution interstellar \sodii\
spectra of BD$+$56\arcdeg 578 and BD$+$56\arcdeg 571, separated by
$\sim$ 2\farcm 5, are quite similar (see Figure~2a); the most apparent
difference is slightly more interstellar absorption at intermediate
velocities toward BD$+$56\arcdeg 571.  It is difficult to discern from
the moderate resolution data if this subtle variation is the result of
a change in the relative intensity and/or velocity of the intermediate
velocity gas in front of these stars or if the variation is indicative
of an additional intermediate velocity component that is present
toward BD$+$56\arcdeg 571, but not toward BD$+$56\arcdeg 578.  The
high-resolution Coud\'e Feed spectra (see Figure~3) clearly reveal 
the presence of an extra intermediate velocity absorption component 
toward BD$+$56\arcdeg 571.  Thus, the moderate resolution ($\sim 
12$~\kms) spectra provide a gross description of interstellar \sodii\ 
absorption toward the double open cluster, but the higher resolution 
spectra are necessary to understand the details of the interstellar
absorption.

The high-resolution spectra shown in Figure~3 do not reveal any
apparent patterns in the interstellar \sodii\ absorption toward h and
\xper.  In general, the high-resolution spectra reveal a multitude of
interstellar absorption components along any given sightline and no
two high-resolution spectra are identical.  Toward \xper, eight of the
ten sightlines lie within a $5'$ ($\sim 3$~pc) radius of the cluster
core.  These sightlines have angular separations that range from $\sim
0\farcm 5$ ($\sim 0.3$~pc) to $8\farcm 5$ ($\sim 5$~pc), with an
average angular separation of $\sim 3\farcm 2$ ($\sim 2$~pc).  Even
these closely-spaced sightlines overwhelmingly show the chaotic
interstellar absorption structure in the Perseus arm gas with
individual components varying in velocity and strength on the smallest
angular scales probed.  For example, BD$+$56\arcdeg 574 and
BD$+$56\arcdeg 575 have an angular separation of $0\farcm 75$ ($\sim
0.4$~pc at the distance to \xper), yet the interstellar absorption of
intermediate velocity Perseus arm gas toward BD$+$56\arcdeg 574 is
much more pronounced.  In contrast to \xper, only four of the fourteen
sightlines toward \hper\ are projected within a $5'$ ($\sim 3$~pc)
radius of the cluster core with angular separations ranging from $\sim
0\farcm 6$ ($\sim 0.35$~pc) to $4\farcm 1$ ($\sim 2.5$~pc) and an
average angular separation of $\sim 2\farcm 5$ ($\sim 1.5$~pc).  Of
these four stars, HD\,14134 and BD$+$56\arcdeg 524 have the smallest
angular separation of $0\farcm 6$ ($\sim 0.35$~pc at the distance to
\hper).  The moderate resolution spectra toward these two stars are
vaguely similar and show fairly strong interstellar \sodii\ absorption
in the intermediate and higher velocity Perseus arm gas.  The
high-resolution spectra, however, show dramatic differences.  The
interstellar \sodii\ absorption spectrum toward HD\,14134 shows two
saturated components at $V_{Hel} \sim -54$ and $-66$~\kms\ as well as
a fairly well-defined gap between the Perseus and Orion arm gas.  The
interstellar \sodii\ absorption spectrum toward BD$+$56\arcdeg 524, on
the other hand, shows relatively strong absorption at all velocities
associated with Perseus arm gas such that no gap is apparent between
it and the Orion arm gas.

A comparison between the interstellar \sodii\ absorption spectra
toward \xper\ with respect to the spectra toward \hper\ does not
reveal many obvious trends.  It appears as if the interstellar \sodii\
absorption toward \hper\ is generally more saturated than toward
\xper.  To further examine these spectra for any discernible patterns
in the interstellar \sodii\ absorption, we created gray-scale images
of the high-resolution spectra and arranged them spatially from west
to east as shown in Figure~4.  In this figure, we also indicate the
average heliocentric velocities of h and \xper\ at $-44.8$ and
$-42.5$~\kms\ (Liu, Janes, \& Bania 1989), respectively.  In this
image, strong \sodii\ absorption is black, more moderate absorption is
gray, and low absorption is white.  These absorption line images
confirm the impression that there is more interstellar \sodii\
absorption toward \hper.  The most noticeable feature in the spectral
line images shown in Figure~4 is the presence of enhanced interstellar
\sodii\ absorption toward \hper\ at $V_{Hel} \sim -20$~\kms\ that is
not seen toward \xper.  We discuss the nature of this $-20$~\kms\
absorption in more detail in \S4.3.  In addition to the $-20$~\kms\ 
component that is only detected toward \hper, we also notice a component 
that in present in most, if not all, of our high-resolution spectra 
between $-50$ and $-60$~\kms.  This velocity component may be indicative 
of a large-scale interstellar structure toward h and \xper\ associated
with the Perseus spiral arm.

\section{Discussion}

Many observations have been reported that detail the structure seen in
interstellar \sod\ absorption toward high proper motion stars,
binaries, open clusters, and globular clusters.  In this section, we
compare the interstellar \sod\ absorption toward h and \xper\ with
observations toward other open clusters and OB associations where it
may be possible to use multi-object spectroscopy to investigate
interstellar absorption toward a large number of sightlines.  These
comparisons may allow us to discern common properties of interstellar
\sod\ absorption toward clusters of early-type stars.  Then, we
investigate the question of what characteristics define an
interstellar cloud.  Finally, we discuss the nature and origin of the
$-20$~\kms\ interstellar \sodii\ absorption component that is detected
toward \hper, but not toward \xper.

\subsection{Interstellar \sod\ Absorption Toward Open Clusters and 
OB Associations}

Among the first ``stationary'' or interstellar lines reported (Heger
1919), \sodd\ observations have a long history of being used to probe 
the structure of the ISM.  Furthermore, the interstellar \sodd\ absorption
lines are among the strongest optical interstellar absorption lines 
typically detected.  Thus, we begin our discussion of the complex
interstellar structure toward h and \xper\ by examining past studies
of interstellar \sodd\ absorption toward open clusters and OB
associations.  Through this, we can examine whether the velocity and
intensity variations in interstellar \sodd\ absorption toward h and
\xper\ are commonly seen toward clusters of early-type stars and
how this affects our understanding of the ISM in the Galaxy.

\subsubsection{h and \xper\ (NGC 869 and NGC 884)}

As mentioned previously, moderate resolution ($\sim 9$~\kms)
interstellar \caii\ and \sod\ absorption observations toward 112 stars
in the northern Milky Way were obtained by M\"unch (1953; 1957) as
part of a larger program to determine if the interstellar lines in
distant stars might reveal the large-scale structure of interstellar
gas in the Galaxy.  Of these stars, 16 are members of the Perseus OB1
association and are projected toward h and \xper.  M\"unch (1953;
1957) determined that the interstellar \sodd\ absorption line spectra
toward h and \xper\ showed two distinct components: (1) a blue-shifted
``V'' component associated with interstellar gas in the Perseus spiral
arm and (2) a highly saturated ``R'' component associated with the
local Orion spiral arm gas.  In particular, M\"unch (1953) noted that
the shape of the line profiles of the Perseus arm gas suggested that
this ``V'' component resulted from the superposition of a number of
fainter narrow lines and that the the shape of the ``V'' component
varied widely from star to star in both velocity and intensity.  The
detailed analysis of these data shows that some of the interstellar
\sod\ absorption in the ``V'' component with radial velocities that
are comparable to those of the background stars and a component in the
Perseus arm gas that is systematically shifted to the violet with
respect to stars by $\approx 20$~\kms.  As seen in Figure~4, this
component is detected in our high-resolution KPNO Coud\'e Feed spectra
toward h and \xper\ as an interstellar \sodd\ absorption feature at
heliocentric velocities between $-50$ and $-60$~\kms.  M\"unch (1957)
has suggested that this blue-shifted velocity component could be
indicative of a coherent expansion of interstellar gas around the OB
associations and open clusters in the Perseus spiral arm.

In order to better understand the nature of the \sod\ absorption
toward the Perseus arm, M\"unch (1957) compared the interstellar
\sodd\ profiles with \hi\ 21-cm emission line data.  M\"unch (1957) 
determined that the peaks seen in 21-cm emission had similar velocities 
to the troughs seen in \sod\ absorption, leading him to conclude that 
the observed \sod\ lines must predominantly arise in \hi\ regions.  
Furthermore, M\"unch (1957) noted that the optical absorption line 
profiles show more variations on smaller scales than the 21-cm
emission line profiles because the beam size of the radio observations
($\sim 2\fdeg 5$) samples many more interstellar clouds along the 
line-of-sight than an individual interstellar absorption sightline.
Thus, the \hi\ data may be able to provide information on the large-scale
structure of the ISM and absorption line studies are better suited
to examine small-scale variations.

\subsubsection{Carina Nebula (NGC~3372)}

As much as h and \xper\ are among the most studied nearby Galactic
clusters accessible to northern telescopes, the Carina Nebula
(NGC~3372) and its associated open clusters Trumpler~16 and
Collinder~228 (hereafter Tr~16 and Cr~228, respectively) are among the
most studied southern objects.  The discovery of complex interstellar
absorption line structure toward the Carina Nebula was first reported
by Walborn \& Hesser (1975) on the basis of \caii\ observations.
Subsequent \sod\ observations toward stars in Tr~16 and Cr~228
(Whiteoak \& Gardner 1980; Walborn 1982; Garc\'{\i}a \& Walborn 2000)
have revealed complex interstellar \sodd\ absorption toward the Carina
Nebula in which none of the line profiles are identical.

Whiteoak \& Gardner (1980) investigated the interstellar \sodd\ 
absorption toward nine early-type stars that are associated with
or behind the Carina Nebula using the Coud\'e spectrograph on the 
Mt. Stromlo 74-in telescope with a velocity resolution of $\sim 
4$~\kms.  The nine stars in their survey have angular separations 
ranging from 1$'$ to $30\farcm 5$, which corresponds to projected 
physical separations of 0.6 to 21.5~pc at a distance to Carina of 
2.2~kpc (Walborn 1995).  These interstellar \sod\ absorption spectra 
reveal a wide saturated feature centered on $\sim -5$~\kms\ which
analogous to the Orion arm gas component that is seen toward h
and \xper.  Another noticeable feature in these spectra is a
prominent absorption component near $-30$~\kms\ that may be 
associated with H$_2$CO (Gardner, Dickel, \& Whiteoak 1973) and
OH (Dickel \& Wall 1974) absorption detected at $-25$~\kms.

Walborn (1982) examined interstellar \sod\ absorption toward 22 
stars within the Carina nebula using the CTIO 4-m telescope and
echelle spectrograph with a velocity resolution $\sim 7$~\kms.  
The stars in Walborn's (1982) investigation, which included 8 of 
the 9 stars surveyed by Whiteoak \& Gardner (1980), have angular 
separations ranging from $0\farcm 23$ to $44\farcm 4$ which 
corresponds to projected physical separations of $\sim 0.15$ to 
26.6~pc at the distance to Carina.  Using this somewhat larger
number of sightlines, Walborn (1982) concluded that the interstellar
\sodd\ absorption at intermediate velocities ($-40 \simless V_{Hel}
\simless -20$ ~\kms) could be interpreted as arising in the front 
side of an expanding shell because these velocity components are
similar to those seen in nebular emission lines.  The higher velocity 
($V_{Hel} < -40$~\kms) interstellar \sodd\ absorption line components, 
however, show relatively little systematic structure and most are 
different from sightline to sightline.  Based on the rapid spatial 
variations in interstellar absorption between closely separated 
stars, Walborn (1982) concluded that the variations must be formed 
relatively near the stars.

Garc\'{\i}a \& Walborn (2000) continued the investigation of
interstellar structure in the Carina nebula with \caii\ and \sod\
observations of an additional 24 stars in Tr~16 and Cr~228 with a
velocity resolution of $\sim 24$~\kms.  They also report that the
structure of the interstellar absorption line profiles change on very
small angular scales that may be attributed to small differences in
distance to the stars.  Furthermore, Garc\'{\i}a \& Walborn (2000)
find that there are fewer higher velocity components toward fainter,
less massive cluster members and suggest that the higher velocity
components are produced in interactions between the stellar winds and
the ambient medium.

\subsubsection{Ara OB1}

Whiteoak \& Gardner (1985) investigated the kinematics of 
interstellar gas toward a molecular cloud/\hii\ region at 
$l = 336\arcdeg$, $b = -1\fdeg 5$.  In addition to obtaining 
4.8~GHz H$_2$CO absorption observed against 5~GHz continuum 
emission, they obtained high-resolution ($\sim 4$~\kms) 
interstellar stellar \sod\ observations toward 28 stars in 
the stellar association Ara OB1 ($335\arcdeg < l < 340\arcdeg$, 
$-2\arcdeg < b < 1\arcdeg$).  The stars in this study have 
angular separations ranging from $1\farcm 6$ to $\sim 275'$ 
which corresponds to projected physical separations of 0.65 
to $\sim 110$~pc at a distance of 1.4~kpc to Ara OB1.

The interstellar \sodd\ absorption line profiles toward these
stars reveal a complex interstellar structure; all of the 
line profiles are different.  Most of the \sodd\ absorption
profiles show several features that are highly saturated and
blended together.  The optical depths in the individual components 
are high ($\tau \sim 5$) with column densities, $N$(\sod) $\sim 5 
\times 10^{12}$~cm$^{-2}$, that are typical of dense clouds 
(Whiteoak \& Gardner 1985).  The interstellar \sod\ absorption 
line spectra show three predominant groupings of components at 
heliocentric velocities of $\sim -40$, $-20$, and $0$~\kms\ that
appear to be related to large-scale galactic features, such as
the local Orion arm at $0$~\kms\ and the Carina-Sagittarius
spiral arm at $-25$~\kms\ (Georgelin \& Georgelin 1976). 

\subsubsection{Sco OB1}

In addition to noting that the interstellar absorption profiles toward
open clusters and OB associations were complex, M\"unch (1957) also
reported the detection of gas with unusually high negative velocities
that could be indicative of large scale expansion.  This work has
inspired numerous higher resolution investigations of interstellar
absorption toward clusters and OB associations in the Galaxy.  The Sco
OB1 association, at a distance of 1900~pc (Humphreys 1978) and
containing many luminous massive stars, provides an excellent
opportunity to probe interstellar absorption line structure over a
range of physical scales.

Crawford, Barlow, \& Blades (1989) investigated the interstellar \sod\
and \caii\ absorption lines toward 22 stars in and around Sco OB1 at
high spectral resolution ($\sim 1.5$ -- 3.0~\kms).  The stars in their
sample included 13 association members, eight nearby stars, and one
foreground star that range in angular separation from $1'$ to 16\fdeg
5, corresponding to physical distances of $\sim 0.5$ to 500~pc
assuming a 1900~pc distance to the association.  The interstellar
\sodd\ absorption profiles toward the distant stars in this sample are
complex and resemble the high-resolution interstellar \sodd\
absorption profiles toward h and \xper.  In fact, the interstellar
\sodd\ absorption spectra toward the two most closely-spaced stars,
HD~152233 and HD~152234, show dramatically different velocity profiles
(see Figures~2d and 3c of Crawford et al. 1989).  The \sodd\ spectra
show complex interstellar \sodd\ absorption profiles at at velocities
less than $-20$~\kms, saturated interstellar \sodd\ absorption
heliocentric velocities $-20 < V_{Hel} < 0$~\kms, and absorption
components at positive velocities ($V_{Hel} > +9$~\kms) in four of the
interstellar \sodd\ absorption spectra.  The interstellar gas with
velocities $V_{Hel} \le -20$~\kms\ is interpreted as arising from
absorption in the Sagittarius spiral arm with the most negative
velocity components representing the front side of an expanding shell
around Sco OB1.  The saturated interstellar absorption detected at
velocities $-20 < V_{Hel} < 0$~\kms\ is most likely caused by
absorption from local, foreground clouds.  The positive velocity
components that were only detected in four of the \sodd\ spectra may
represent either the receding parts of an expanding shell or
low-density intercloud material.

Crawford (2001) obtained ultra-high resolution ($\sim 0.35$~\kms)
interstellar \sod\ and K\,{\sc i} absorption profiles toward 3 of the
sightlines investigated by Crawford et al. (1989).  These ultra-high
resolution spectra show even more remarkable structure in the
interstellar \sodd\ absorption line profiles than were revealed in the
high resolution data of Crawford et al. (1989).  In general, the ultra-high
resolution spectra show approximately twice the number of interstellar
absorption line components than seen the high-resolution data.

\subsubsection{The Pleiades (M~45)}

White at al. (2001) investigate the interstellar \sodd\ absorption
toward 36 stars in and around the Pleiades to better understand the
kinematics of interstellar gas as it interacts with the Pleiades at
high spectral resolution ($\sim 1.5$~\kms) and at high
signal-to-noise.  The sightlines in this sample had angular
separations ranging from $\sim 2'$ to 8\arcdeg\ which corresponds to
physical scales from $\sim 0.07$ to 16~pc, respectively, at a distance
to the Pleiades of 118~pc (van Leeuwen 1999).  In general, these
interstellar \sodd\ absorption spectra have simpler line profiles than
those seen toward more distant clusters and OB associations with the vast
majority of spectra (75\%) only showing one or two velocity components,
typically at $V_{Hel} \sim 15 \pm 2$~\kms.

Although most of the spectra in this investigation have relatively
simple absorption profiles, the spectra do show some complexity.  
The two sightlines with the smallest angular separation ($\sim 2'$;
0.07~pc), toward the Pleiades members HD~23629 and HD~23630, show 
two and three absorption components, respectively.  This additional 
velocity component detected toward HD~23630 at $V_{Hel} = +5.6$~\kms\
may be indicative of a small scale physical structure in the ISM.  
The next closest sightline to HD~23630 is toward cluster member HD~23642 
which lies $\sim 11'$~N ($\sim 0.37$~pc) and shows only one absorption
component at $V_{Hel} = +15.8$~\kms, further indicating that the
component at $V_{Hel} = +5.6$~\kms\ covers only a small region on the 
sky.  Those sightlines which show more than two velocity components
(25\% of the sample) are randomly distributed across the region surveyed
and include sightlines to both member and non-member cluster stars.
Furthermore, the number of interstellar absorption components detected
does not depend on the distance to the background star.  The spectra 
toward the two most distant stars in the White et al. (2001) sample, 
HD~25201 ($d = 332$~pc) and HD~26128 ($d = 245$~pc), show three and one 
absorption component, respectively.  

\subsubsection{Orion}

Price et al. (2001b) obtained ultra-high resolution ($\sim 0.35$~\kms)
interstellar \sod, Ca\,{\sc ii}, K\,{\sc i}, CH, and CH$^{+}$
absorption spectra toward 12 stars in Orion, including four stars in
M\,42.  The individual absorption components detected toward each
sightline were then used to examine the kinematics of the Orion region
as well as that of the ISM.  Because the stars in this sample lie at
distances ranging from $\sim 250$ to $450$~pc and have angular
separations ranging from 0\farcm 2 -- 20\fdeg 25 ($\sim 0.03$ --
$160$~pc, assuming a common distance of 450~pc), it is possible to
examine both the small and medium scale structure of the intervening
material.

These ultra-high resolution data have revealed the presence of up to
16 absorption components per sightline which more than doubles the
number of interstellar components previously reported along these
sightlines (e.g., Hobbs 1978; O'Dell et al. 1993; Welty, Morton, \&
Hobbs 1996).  The detailed comparison of absorption components
detected in \sod, Ca\,{\sc ii}, and K\,{\sc i} were used to identify
individual interstellar clouds; clouds with similar physical
parameters were then grouped into absorption systems.  The
observations along sightlines with small angular separations (e.g.,
toward M\,42 and $\lambda$~Ori) have revealed variations in the
absorption line profiles that imply the presence of structure down to
0.03~pc.  Furthermore, comparison of the interstellar \sod\ absorption
spectrum toward $\zeta$~Ori obtained by Price et al. (2001b) with one
obtained by Welty, Hobbs, \& Kulkarni (1994) shows evidence for
temporal variations in the line profile over a 6~yr time period,
indicating structure in the ISM down to scales of $\approx 7$~au,
given a proper motion of $\zeta$~Ori of 4.73~mas~yr$^{-1}$ (ESA 1997).
This temporal variation in the line profile is similar to that which
has been detected toward $\delta$~Ori (Price, Crawford, \& Barlow
2000; Price et al. 2001a) and toward HD~32039/40 (Lauroesch, Meyer, \&
Blades 2000).  Even though temporal variations are detected in the
interstellar \sod\ absorption profiles, they are not seen in the
interstellar Ca\,{\sc ii} absorption along this sightline.

\subsubsection{Comparison With Our Results}

The previous studies, as well as our current investigation of
interstellar \sodd\ absorption toward early-type stars reveal complex,
multiple-component line profiles.  All of the sightlines toward any
particular open cluster or OB association show different interstellar
\sodd\ absorption line profiles.  The differences in the absorption
line profiles are most apparent for gas that is located in other
spiral arms of the Galaxy, however, nearby sightlines toward the
Pleiades and stars in Orion also show considerable variations in the
interstellar \sodd\ profiles.  Furthermore, significant differences
among the line profiles toward stars in each cluster can be seen down
to the minimum projected stellar separations of $\sim 0\farcs 1$~pc
(e.g., see Figures~2 \& 3).  Thus, it would seem that the variations
observed in interstellar \sodd\ absorption toward stars in open
clusters and OB associations are a relatively common phenomenon.  As
previously discussed, it remains uncertain if the differences in the
interstellar \sodd\ absorption line profiles are tracing density variations
in interstellar gas, fluctuations in the ionization balance, the
geometry of interstellar material, or a combination of these and
other effects.

\subsection{What Is A Cloud?}

%\subsubsection{An Historical Perspective}

The first interstellar absorption line reported was detected 
toward the spectroscopic binary $\delta$~Ori by Hartmann (1904), 
who noted that the \caiik\ line did not share in the periodic 
variations of the stellar lines but remained ``stationary'' 
throughout the observations.  Subsequently, stationary lines of 
\caiik\ were detected in the spectra of other stars such as 
$\rho$~Leo (Harper 1914).  This phenomenon was extended by Heger 
(1919) when stationary lines of \sodi\ and D$_2$ were discovered 
in the spectra of $\delta$~Ori and $\beta$~Sco.  Heger (1919) also
noted that the velocities of the stationary \sodd\ lines in these 
spectra matched the previously measured velocities of stationary 
\caiih\ and K lines toward these stars within the observational
errors.

The origin of the stationary lines remained in dispute for
approximately 25 years after their discovery.  The two
predominant theories to explain the nature of the stationary 
lines were that (1) the lines were due to absorption of the 
background stellar continuum by an intervening cloud of calcium 
and/or sodium in space or (2) that the lines were due to a 
calcium and/or sodium envelope that surrounded the stellar 
source.  Until 1923, the stationary lines had only been detected 
toward high temperature spectroscopic binaries\footnote{It 
should be noted that $\rho$~Leo is not a spectroscopic binary.  
Harper (1914) considered $\rho$~Leo to be a spectroscopic binary 
based on a report by Campbell \& Albrecht (1909) even though he 
was unable to determine a period for the star from his own data.} 
with spectral types between B0 and B3.  Based on this evidence, 
Young (1920) concluded that the stationary lines are most likely 
produced in a low density gas that was associated with the binary 
systems.  Plaskett (1923) challenged this conclusion, however,
when his investigation of the spectra of $\sim 40$ O-type and 
Wolf-Rayet stars revealed that stationary lines of \caiih\ and K 
were present toward all of the stars in his sample regardless if 
the star were a binary or singular.  Furthermore, Plaskett (1923) 
found that the velocities of the stationary H and K lines differed 
considerably from the velocities of stellar lines in the spectra by 
as much as $60$~\kms.  Therefore, Plaskett (1923) concluded that 
diffuse gaseous matter containing ionized calcium (and neutral 
sodium) arise in a fairly widespread cloud of tenuous material that 
is not local to the background star, but was interstellar gas.

The physical conditions of diffuse interstellar gas was first
investigated by Eddington (1926).  Using general physical
considerations, Eddington (1926) estimated that the diffuse
interstellar gas had a density of $\sim 10^{-24}$~g~cm$^{-3}$, a
temperature of $\sim 10^4$~K, and was uniformly distributed throughout
the Galaxy.  Thus, interstellar absorption lines should be present in
the spectra of stars of all classes and the strength of the absorption
should increase with distance to the background star.  Because the
detection of interstellar absorption in the spectra of late-type stars
is problematic due to the increasing strength and narrow width of
stellar absorption lines, most investigations of interstellar
absorption focused on the spatial distribution of the gas toward
early-type stars.  Struve (1928) systematically measured the intensity
of interstellar \caiik\ toward 1718 early-type stars that ranged in
apparent magnitude by 10.5~mag and found that the line strength
linearly increased with apparent magnitude and presumably with
distance.  In order to determine the relative distances between the
star and the center of absorption in a cloud by using the theory of
differential Galactic rotation (Lindblad 1927), Plaskett \& Pearce
(1930) measured the stellar and interstellar \caiik\ line velocities
toward $\sim 250$~OB stars.  Plaskett \& Pearce (1930) found that the
galactic rotational term of stars was approximately twice the galactic
rotational term of the gas and concluded that the absorbing gas was
uniformly distributed.  Photometric measurements of interstellar line
intensities led to the conclusion that the interstellar gas was not
homogeneous, but that the observed line profiles could be accounted
for by making the hypothesis that interstellar gas occurs in discrete
clouds which participate in the general galactic rotation in addition
to having significant random motions (Wilson \& Merrill 1937).

The discovery of the ISM through absorption line spectroscopy led to
what can be described as a ``standard cloud'' model in which each
velocity component comprising an interstellar absorption line profile
in stellar spectra corresponds to an individual cloud with a radius of
a few parsecs (Spitzer 1978).  Statistically, about five of these
standard clouds are detected per kiloparsec along any particular
sightline (Hobbs 1974).  It is important to note, however, that
advances in spectral resolution have radically altered this simple
picture.  As higher spectral resolutions are achieved, the number of
discrete absorption components along any particular line-of-sight
tends to increase.  Beals (1936) first reported the ``irregularity of
structure'' in the interstellar absorption profiles toward stars in
Orion (e.g., $\delta$~Ori and $\zeta$~Ori ) that indicated the presence 
of two interstellar clouds.  As discussed in \S 4.1.6, the ultra-high 
resolution observations of interstellar absorption toward stars in Orion 
(Price et al. 2001b) reveal up to 10 and 12 interstellar clouds toward
$\delta$~Ori and $\zeta$~Ori, respectively.  Thus, the ``standard cloud''
model of the ISM which was constructed on the interpretation of
moderate resolution interstellar absorption spectra cannot adequately 
describe the ISM as viewed at ultra-high resolution.  Consequently,
the definition of an interstellar cloud strongly depends upon
the resolution with which it is observed.

The view of the ISM as determined from interstellar absorption is
sharply contrasted with the appearance of the ISM as determined from
emission line imaging studies which reveals the ISM to be comprised of
loops, filaments, shells, and sheets of gas.  Thus, the relationship
between interstellar structures seen in emission and interstellar
clouds inferred from absorption is not clear.  A fundamental problem
in reconciling the differences between these views of the ISM is
pencil-beam nature of interstellar absorption line investigations and
the larger beam sizes of \hi\ 21-cm emission line surveys.  One
approach in attacking this problem is to map interstellar absorption
lines of intervening gas toward open clusters and OB associations 
by obtaining many spectra along sightlines which probe a range of
angular separations through the use of multi-object spectroscopy.

We have seen in \S 4.1 that observations of interstellar \sod\ toward
stars in open clusters (e.g., h \& \xper, Tr~16 \& Cr~228, \& the
Pleiades) and OB associations (e.g., Ara~OB1 \& Sco~OB1) reveal
complex absorption line profiles.  As shown in Figure~2, all of the
sightlines toward the 150 stars in the vicinity of h and \xper\ for
which we have obtained WIYN Hydra moderate resolution spectra exhibit
different \sodd\ line profiles for every sightline.  The variations in
the absorption line profiles are most easily seen in optically-thin
Perseus spiral arm gas with velocities between $-75$ and $-20$~\kms.
Furthermore, these variations can be seen down to the minimum angular
separation of $\sim 11\farcs 5$ which corresponds to a projected
physical separation of $\sim 0.1$~pc at the distance to the Perseus
gas.  As shown in Figure~3, the \sodd\ profile variations are even
more dramatic in our high-resolution KPNO Coud\'e Feed spectra toward
24 of the h and \xper\ stars which have angular separations as small
as $\sim 30''$ ($\sim 0.3$~pc).

In the ``standard'' cloud model, each of the interstellar \sod\
absorption components seen toward h and \xper\ represents an
individual, spherical, isothermal cloud with a diameter of $\sim 
5$~pc (Spitzer 1978).  Thus, at the 2~kpc distance to the Perseus 
spiral arm, the minimum angular size of a ``standard'' interstellar 
cloud along the line-of-sight would be $\sim 8\farcm 6$.  As seen 
in Figure~3, along any given sightline toward h and \xper\ for 
which we have obtained high-resolution interstellar \sod\ absorption 
spectra, the approximate number of components it would take to 
adequately fit the line profile is $\sim 10$ for the gas in the 
Perseus arm ($-75 \le V_{Hel} \le -20$~\kms).  Therefore, the 
assumption that each detected interstellar \sod\ absorption component 
arises from within an individual cloud gives rise to a complicated 
situation where the ISM of of the Perseus spiral arm would appear 
to consist of hundreds of individual clouds.

Given the size of a ``standard'' cloud and its projected angular
diameter at the distance of h and \xper, it is not unreasonable to
assume that some fraction of the absorption lines that are detected
along different sightlines arise in the same ``cloud''.  Some of the
sightlines with the smallest angular separations between them have
fairly distinct interstellar \sod\ absorption spectra.  For example,
HD~14443 and BD$+$56\arcdeg 571 toward \xper\ have an angular
separation of $\sim 0\farcm 5$.  Of the five easily discernible
spectral components in the Perseus arm gas toward HD~14443 (see
Figure~5a), only those components with intensity minima at $-65$,
$-45$, and $-35$~\kms\ appear to directly correspond to absorption
features seen toward BD$+$56\arcdeg 571, i.e., the shift in the 
absorption troughs at these velocities is less than 3.5~\kms.  
HD~14134 and BD$+$56\arcdeg 524 in \hper\ have an angular separation 
of $\sim 0\farcm 6$.  Likewise, as seen above, only three of the five 
obvious absorption components toward HD~14134 at $-70$, $-55$, and 
$-42$~\kms\ appear to directly correspond to features toward 
BD$+$56\arcdeg 524 (see Figure~5b).  Thus, these data show that 
although some of the detected gas is seen toward adjacent stars, 
some is not.

It remains uncertain, however, over what velocity interval we can
legitimately call interstellar \sod\ absorption components toward
different sightlines one cloud.  For example, if we were to analyze
each sightline toward h and \xper\ in isolation, we would need at
least 5 absorption components to adequately fit the interstellar
absorption detected in the Perseus arm gas, implying the presence of
at least that many clouds along the line-of-sight.  If we were to
examine interstellar absorption spectra along two nearby sightlines,
we would observe $\sim 5$ absorption components toward each star, but
2--3 of the components would be seen toward both stars.  This
effectively reduces the number of clouds along the line-of-sight in
the Perseus arm gas from 10 to 7 or 8 toward those two stars.  It is
only when a relatively large number of stars are included in the
analysis that one can begin to discern the larger scale structures in
the ISM toward h and \xper.  For example, we detect interstellar
\sodd\ absorption at a heliocentric velocity of $\sim -55 \pm 5$~\kms\
toward most of the stars for which we have obtained high-resolution
spectra that is suggestive of a large-scale interstellar feature
toward the double cluster that appears to contain medium and small
scale variations within it.  Furthermore, it is possible that the
velocity dispersion of this cloud as a whole is larger than the
velocity dispersion measured for the density peaks within this cloud
that are determined from the individual spectra.  Thus, these
individual density peaks could have heliocentric velocities that
differ by several~\kms, but nevertheless are part of a larger cloud.
The view that some of the differences in interstellar \sodii\
absorption detected among sightlines represents variations within
larger scale absorption systems is not without precedent.  The
ultra-high resolution survey of interstellar absorption toward stars
in Orion (Price et al.  2001b) revealed $\sim 15$ varying velocity
components toward stars in the Trapezium.  Although these ultra-high
resolution spectra reveal some similarities in the interstellar
absorption profiles, they also reveal clear differences.  Despite the
differences detected in the interstellar absorption features, Price et
al. (2001b) argue that the data are best explained if the individual
velocity components comprise part of a larger interstellar absorption
system.  Therefore, based on our moderate and high resolution spectra,
as well as the qualitative comparison to the ultra-high resolution
survey of interstellar absorption toward Orion, the individual
interstellar absorption components detected toward h and \xper\
probably do not indicate the presence of many small spherical clouds,
but rather larger sheets or systems of gas within which we observe
variations in the strength, number, and velocity of components due to
differences in the physical conditions on smaller scales and within
which we detect differences in the gas density.

Because \sod\ is not a dominant species in the ISM, it is usually
necessary to observe other interstellar lines in order to determine if
the interstellar \sod\ absorption line components are tracing the
physical structure or the physical conditions of the gas.  We have
obtained \hi\ 21-cm emission line data toward h and \xper\ from the
Low Resolution DRAO Survey (LRDS) of the Galactic Plane\footnote{The
LRDS of the Galactic Plane was carried out as a part of the Canadian
Galactic Plane Survey project and are publicly available at
http://www.drao-ofr.hia-iha.nrc-cnrc.gc.ca/science/LRDS/survey.html.}
(Higgs \& Tapping 2000) to directly compare the \sodd\ line profiles
with the \hi\ line profiles.  In Figure~6, we show three \hi\ 21-cm
line profiles from the LRDS that are roughly centered on \hper, \xper,
and an area located between these two clusters.  In general, the \hi\
21-cm emission line profiles show smoother, more continuous variations
in the Perseus arm gas when compared to our Coud\'e Feed interstellar
\sodii\ absorption line spectra.  These differences may be due to the
$12'$ beam of the LRDS sampling many individual smaller clouds with
different velocities along the line-of-sight.  The \hi\ 21-cm emission
line profiles show three peaks in the Perseus arm gas at velocities of
$V_{Hel} \sim -65$, $-55$, and $-42$~\kms\ which are similar to some
velocity components detected in interstellar \sod\ absorption.  The
21-cm emission at $V_{Hel} \sim -105$~\kms\ is emission from gas that
lies beyond the Perseus arm.  Of particular interest is the $\sim
-55$~\kms\ 21-cm emission peak because interstellar \sod\ absorption
is detected at velocities between $-60 \le V_{Hel} \le -50$~\kms\ (see
\S 3.3).  We have examined the LRDS \hi\ datacube of the h and \xper\
region and find that an \hi\ feature having an angular diameter of
$\sim 1\arcdeg$ ($\sim 36$~pc) and a velocity of $\sim -55$~\kms\ is
roughly centered on h and \xper.  Therefore, we conclude that the
$V_{Hel} \sim -55 \pm 5$~\kms\ component represents an intervening
interstellar cloud.

In addition to comparing the interstellar \sod\ absorption line
profiles with \hi\ 21-cm emission toward h and \xper, we have also
examined interstellar absorption from the $\lambda 5780$ and $\lambda
5797$ diffuse interstellar bands (DIBs).  Even though not one DIB
carrier has been positively identified since their discovery by Heger
(1922), recent developments indicate that most DIB carriers are
probably large carbon-bearing gas phase molecules such as polycyclic
aromatic hydrocarbons, fullerenes, and/or carbon chains (Salama et
al. 1996; Foing \& Ehrenfruend 1997; McCarthy et al. 2000).  Thus, it
is not unreasonable to assume that the DIBs trace dense gas that could
be indicative of individual clouds toward h and \xper.  Unlike the
interstellar \sod\ profiles, the DIB profiles show little
sightline-to-sightline variations and have velocities typically
associated with local Orion arm and low-velocity Perseus arm gas.
Toward three closely-spaced stars (angular separations $\le 6'$;
projected linear separations $\le 3.5$~pc), however, we have
spectroscopically resolved a $\lambda 5797$ DIB component associated
with higher velocity gas in the Perseus arm.  Our Coud\'e Feed
interstellar absorption spectra toward the two brightest of these
stars, BD$+$56\arcdeg 488 and BD$+$56\arcdeg 508 (see Figure~7), show
an intense, narrow \sod\ absorption component at $V_{Hel} = -70$~\kms\
that is not detected toward any other of the sightlines toward h and
\xper\ for which we have obtained high-resolution spectra.  Because
this strong $-70$~km~s$^{-1}$ \sod\ absorption component and the
Perseus arm $\lambda 5797$ DIB absorption are only detected toward a
small ($\sim 3.5$~pc) region to the northwest of the core of \hper, it
it likely that they trace an isolated interstellar cloud.

Although the above two examples show that some interstellar \sodii\
absorption components seen in Figure~7 appear to trace discrete clouds
in the Perseus arm gas, it is uncertain whether the interstellar \sod\
absorption components at other velocities do.  The variations in the
interstellar \sod\ line profiles at these other velocities may reflect
environmental variations in parameters such as temperature, pressure,
electron density, and/or radiation field.  The physical conditions of
the Perseus arm gas, however, cannot be determined solely from the
interstellar \sod\ absorption lines.  Therefore, we have performed a
cursory examination of the spectral types of stars toward h and \xper\
as cataloged by Slesnick et al. (2002) in order to determine a rough
estimate of the local background radiation field.  As seen in
Figures~3 \& 4, even stars with small angular separations in the core
of h (e.g., HD~14143 and HD~14134) and \xper\ (e.g., BD$+$56\arcdeg
571 and BD$+$56\arcdeg 563) with identical spectral classifications
show different interstellar \sod\ absorption line profiles, indicating
that the local background radiation field is not the primary cause of
the observed line profile variations.  Because these stars are not
likely at the same distance in the Perseus arm, it is also a
possibility that the differences in the interstellar \sod\ line
profiles are caused by distance effects.

Based upon the data we have examined, the most likely distribution of
interstellar material toward h and \xper\ is in the form of several
large-scale structures at velocities of $\sim -72$, $-65$, $-55$,
$-45$, and $-20$~\kms.  The lower velocity gas ($V_{Hel} \ge
-20$~\kms) is primarily associated with more local, Orion spiral arm
gas.  The intermediate velocity gas ($-55 \le V_{Hel} \le -20$~\kms)
detected in these spectra is roughly centered on the mean heliocentric
radial velocities of both h and \xper\ (Liu, Janes, \& Bania 1989).
As such, these absorption components most likely represent gas in the
Perseus spiral arm that is local to the clusters.  Furthermore, the
intermediate velocity interstellar \sodii\ absorption at $V_{Hel} \sim
-55$~\kms\ is blue-shifted with respect to the mean heliocentric
cluster velocities by $\sim 10$~\kms.  As noted by M\"unch (1953;
1957), this blue-shifted component could represent the front side of
an expanding shell of gas around the double cluster.  Velocity
variations among sightlines in this intermediate velocity gas could be
due to the non-spherical geometry of an interstellar shell around h
and \xper, such as the Orion-Eridanus superbubble that is suspected to
produce the variable interstellar absorption detected toward
$\delta$~Ori (Price et al. 2001a).  These largeer scale interstellar
structures have internal velocity dispersions due to turbulence on
small scales (Elmegreen 1999; Cho, Lazarian, \& Vishniac 2002) and
velocities due to bulk motions on larger scales.  Within these
structures, small and medium scale variations in the physical
conditions of the gas could lead to observed differences in the \sod\
column.  Because recombination rates are strongly dependent upon
density (P\'equignot \& Aldrovandi 1986), neutral species such as
\sod\ would preferentially form in these regions of higher density
(Lauroesch \& Meyer 2003 and references therein).  This, coupled with
the velocity differences among these structures, leads to the very
large number of individual \sod\ components detected toward any
particular sightline as well as the differences between sightlines in
terms of both the strength and velocity of interstellar \sod\
absorption.  The higher velocity ($V_{Hel} \le -55$~\kms) interstellar
\sodii\ absorption, especially that which is only detected at $\sim
-70$~\kms\ toward BD$+$56\arcdeg 488 and BD$+$56\arcdeg 508 and also
corresponds to enhanced $\lambda 5797$ DIB absorption, seems to arise
in a canonical interstellar cloud ($\sim 3.6$~pc).

%One other intriguing posibility is that the observed variations 
%in the interstellar \sod\ absorption profiles are due to biased 
%neutral formation in the density peaks of a much larger cloud or 
%clouds (Lauroesch \& Meyer 2003).  Recent models of the effects 
%of magnetohydrodynamical (MHD) turbulence on interstellar clouds 
%have shown that any particular interstellar cloud can contain
%small-scale density perturbations (Elmegreen 1999; Cho, Lazarian, 
%\& Vishniac 2002).  Because recombination rates are strongly 
%dependent upon density (P\'equignot \& Aldrovandi 1986), neutral 
%species such as \sod\ would preferentially form in these regions 
%of higher density.  Thus, the observed variations in the interstellar 
%\sod\ absorption line profiles toward h and \xper\ could arise in 
%overdense regions from within an individual cloud and not from many
%small clouds.

\subsection{The $V_{Hel} \sim -20$~\kms\ Component} 

As discussed in \S 3.3, we have detected an enigmatic interstellar
\sod\ absorption component at $V_{Hel} \sim -20$~\kms\ in the spectra
of stars that are projected toward \hper\ using our Coud\'e Feed data
(see Figure~4).  This absorption feature is not detected in the
Coud\'e Feed spectra of stars projected toward \xper.  In order to
determine the spatial distribution of the $\sim -20$~\kms\ absorption
component, we have re-examined the 150 Hydra interstellar \sod\
absorption spectra toward h and \xper.  This $\sim -20$~\kms\
component is not resolved from the broad (saturated) absorption
centered near $V_{Hel} \sim 0$~\kms.  Therefore, we use the stars in
our Coud\'e Feed sample to estimate the size of the $\sim -20$~\kms\
absorber even though they have a more limited spatial distribution
than the Hydra sample.  From the angular separations between HD~14092
and BD$+$56\arcdeg 488 and between HD~14250 and BD$+$56\arcdeg 485, we
find that the minimum angular extent of the $\sim -20$~\kms\ velocity
component is at least $36'$ $\times$ $20'$.

%These data, however, do not
%have the velocity resolution necessary to separate the $\sim -20$~\kms\ 
%component from the saturated Orion arm gas at $\sim -10$~\kms.  

The large angular size of the $\sim -20$~\kms\ interstellar \sod\
absorption component implies that the gas in which this feature
appears lies at some distance in front of \hper.  If the gas were
relatively close to the cluster, we would expect to see some
differences in the line profiles due to the ionizing radiation field
from the massive stars in \hper.  Thus, the most simple explanation is
that the $\sim -20$~\kms\ component arises in a discrete neutral cloud
at an unknown distance in front of \hper.  To confirm this hypothesis,
we have examined the 21-cm LRDS datacube at heliocentric velocities
between $-30$ and $-10$~\kms.  This is an ideal method to search for
an isolated cloud on large angular scales because these data
simultaneously provide kinematic and spatial information on the \hi\
distribution.  We find a 21-cm \hi\ emission component at $\sim
-15$~\kms\ that is projected toward \hper, but that is absent toward
\xper, strengthening our conclusion that the $\sim -20$~\kms\
interstellar \sod\ absorption feature arises in a gas cloud that lies
in front of \hper.  It is interesting to note, however, that an
investigation of the LRDS \hi\ 21-cm emission line profiles toward the
cores of h and \xper, where the majority of stars observed with the
Coud\'e Feed are located, does not show any remarkable enhancements in
the neutral gas at this velocity (see Figure~6).  This difference,
or lack there-of, in the individual \hi\ 21-cm emission line profiles
may be the result of self-absorption of neutral hydrogen in the local
gas that affects the shape of the 21-cm line profiles in this velocity
range.

\section{Summary}

We present moderate resolution (12~\kms) interstellar \sod\ absorption
line spectra along 150 sightlines toward the double open cluster h and
\xper\ that were obtained with the WIYN Hydra multi-object
spectrograph.  Additionally, we have used the KPNO Coud\'e Feed
spectrograph to obtain high-resolution (3~\kms) interstellar \sod\
absorption line spectra toward 24 of the brightest stars along the
line-of-sight to h and \xper.  These data have been used to examine
the interstellar \sod\ absorption toward h and \xper\ on angular
scales from $\sim 10''$ to $\sim 5000''$ which correspond to linear
scales of 0.1~pc to 50~pc at the double cluster distance of $\approx
2000$~pc.  Both the moderate and high-resolution spectra reveal
complex interstellar \sod\ absorption toward h and \xper.  A study of
these spectra leads us to the following conclusions:

(1) Multi-object spectroscopy of interstellar absorption toward open
clusters is a powerful, but under-utilized tool to examine the
structure of the ISM.  Studies of interstellar absorption and its
variations from sightline to sightline have historically been limited
to obtaining one spectrum at a time, making it difficult and time
consuming to investigate the structure of the ISM over a range of
scale lengths.

(2) The moderate resolution WIYN Hydra data are remarkable because
they show that no two interstellar \sod\ absorption line profiles are
alike over the range of scale lengths probed.  The high-resolution
KPNO Coud\'e Feed spectra reveal even more complexity in the Perseus
arm gas than the moderate resolution Hydra spectra with the number,
velocity, and strength of individual components changing from
sightline to sightline.  These results are similar to those obtained
by previous investigations of interstellar \sod\ absorption toward
open clusters and OB associations in the Galaxy.  Therefore, it seems
that the variations which are detected in interstellar \sod\ absorption 
line profiles are relatively common.

(3) The ``standard'' cloud model of the ISM in which every individual
absorption line component toward each sightline represents an
individual interstellar cloud is not applicable to these data.  If
each interstellar \sod\ absorption component detected in our Coud\'e
Feed dataset represents an individual interstellar cloud, the Perseus
arm gas toward h and \xper\ would appear to consist of many small
clouds.  Given the finite size of an individual interstellar cloud,
the number distribution of clouds along the line-of-sight to h and
\xper, and the range of angular separations of our sightlines, it
seems probable that some sightlines should intersect the same parcel
of gas.  The difficultly lies in determining over what velocity
interval and change in intensity it is appropriate to call
interstellar \sod\ absorption components toward different sightlines
one cloud.  It is distinctly possible that the ISM in the Perseus arm
gas is comprised of sheets of gas that have large internal velocity
dispersions.  Fluctuations in the density within these sheets of gas
could give rise to the variations observed in our interstellar
absorption line data.  Ultra-high resolution interstellar absorption
surveys toward Orion reveal extremely complex line profiles toward any
individual star (Price et al 2001b).  Although numerous velocity
components are detected toward stars in the Orion sample, Price et
al. (2001b) are able to group absorption components into larger 
absorption systems based upon similarities in velocity and in
Na$^0$/Ca$^+$ ratios.

(4) Because \sod\ is not a dominant species in the ISM, it is
uncertain if the observed variations in the interstellar absorption
line profiles trace changes in the physical structure or the physical
conditions of the gas.  Therefore, other interstellar lines need to be
observed, such as Ca\,{\sc ii}, K\,{\sc i}, or diagnostic UV lines.
Although our moderate and high resolution data do not cover these more
traditional diagnostic lines, we have also examined the interstellar
absorption from the $\lambda 5780$ and $\lambda 5797$ DIBs.  These
DIBs most likely trace dense gas that could be indicative of
individual clouds.  Toward three closely spaced stars, we have
spectroscopically resolved a $\lambda 5797$ DIB component in the
Perseus spiral arm.  Our high-resolution interstellar \sod\ absorption
spectra toward the two brightest of these stars reveal a nearly
saturated, narrow absorption component at $V_{Hel} = -70$~\kms.  It is
likely that this strong \sod\ component traces an individual
interstellar cloud.

(5) The comparison between the high-resolution spectra toward
\xper\ and \hper\ does not reveal many obvious differences between 
the two clusters.  In general, the interstellar \sod\ absorption
toward \hper\ seems more pronounced.  Furthermore, inspection of
spectral images of the high-resolution data reveals a noticeable
interstellar \sod\ absorption component at $-20$~\kms\ toward \hper\
that is not detected toward \xper.  This $-20$~\kms\ interstellar
\sod\ absorption component probably arises in discrete cloud at
an unknown distance.  To verify the physical nature of this feature,
we have examined the LRDS \hi\ 21-cm emission line datacube toward
h and \xper.  We find an \hi\ emission component that correlates in
position and velocity with the $-20$~\kms\ detected in absorption,
confirming our hypothesis that the $-20$~\kms\ absorption component
is associated with an individual cloud.

\acknowledgments

The authors would like to thank the support of the staff of KPNO,
especially Di Harmer and Daryl Willmarth, for their assistance in
obtaining the WIYN Hydra and Coud\'e Feed data presented here.  We also
acknowledge the contributions of D. T. Nguyen for his reductions of
the WIYN Hydra data and S. Cartledge for obtaining some of the of the
Coud\'e Feed spectra presented here.  We would also like to express 
our gratitude to the referee for her/his many helpful comments that
improved the quality of this manuscript.

The Digitized Sky Survey was produced at the Space Telescope Science 
Institute under U.S. Government grant NAG W-2166. The images of these 
surveys are based on photographic data obtained using the Oschin Schmidt 
Telescope on Palomar Mountain and the UK Schmidt Telescope. The plates 
were processed into the present compressed digital form with the permission 
of these institutions.  The LRDS of the Galactic Plane was carried out as 
a part of the Canadian Galactic Plane Survey project and are publicly 
available at: 
http://www.drao-ofr.hia-iha.nrc-cnrc.gc.ca/science/LRDS/survey.html.
The Canadian Galactic Plane Survey is a Canadian project with 
international partners, and is supported by a grant from the Natural 
Sciences and Engineering Research Council of Canada.  Further details 
regarding the CGPS are available from the web site: 
http://www.ras.ucalgary.ca/CGPS. This research has also made use of 
the SIMBAD astronomical database, operated at CDS, Strasbourg, France.

\clearpage

\begin{figure}
\epsscale{0.7}
\figurenum{1}
\plotone{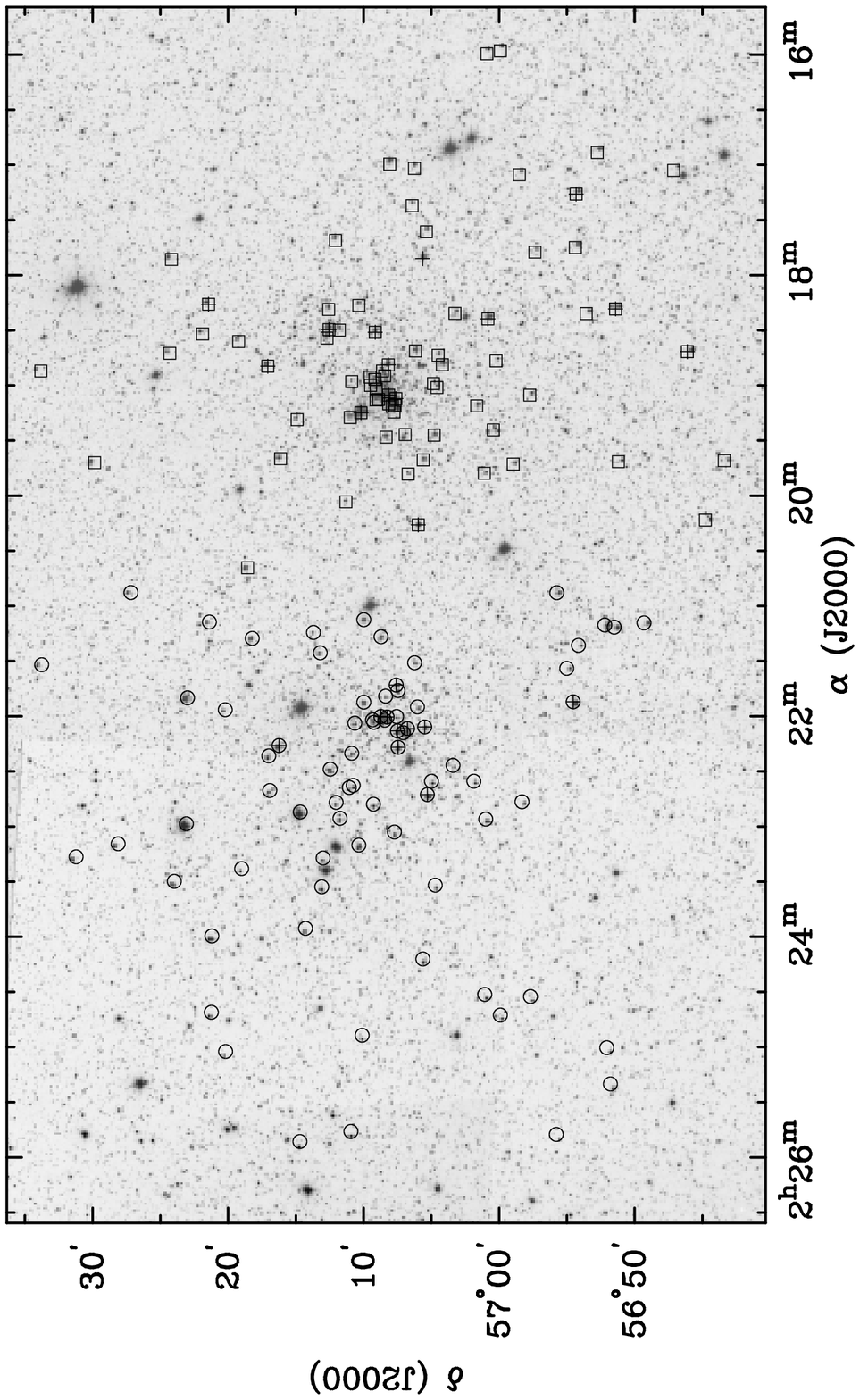}
\caption{DSS mosaicked image of h and $\chi$ Per.  The positions of the
stars toward h and $\chi$~Per for which we have obtained WIYN Hydra 
multi-object spectra of interstellar Na~{\sc i}~D absorption are marked
by boxes and circles, respectively.  The positions of stars for which 
we have obtained KPNO Coud\'e Feed spectra of interstellar Na~{\sc i}~D 
absorption are marked with crosses.}
\end{figure}

\clearpage

\begin{figure}
\epsscale{0.65}
\figurenum{2}
\plotone{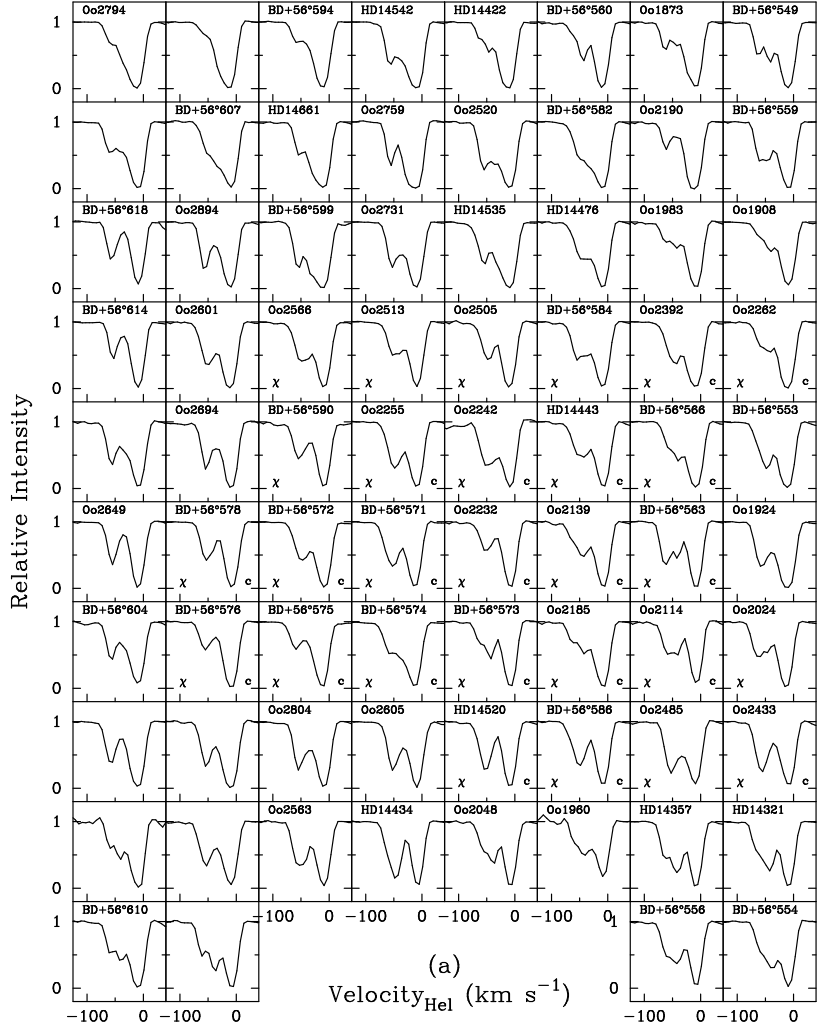}
\caption{{\it (a)\/} Moderate resolution (12~\kms) WIYN Hydra
multi-object spectra of interstellar Na~{\sc i}~D$_2$ (5889\AA)
absorption seen for stars toward $\chi$~Per (stars marked with circles
in Fig.~1).  Spectra toward stars that have been identified as
$\chi$~Per cluster members by Slesnick, Hillenbrand, \& Massey (2002)
have been labeled with a ``$\chi$''.  Where applicable, we have also
included the {\it Henry Draper (HD)\/}, {\it Bonner Durchmusterung
(BD)\/}, and {\it Oosterhoff (Oo)\/} identifications for these stars.
Stars within a $5'$ radius of the cluster core have also been labeled
with a ``c''.  These spectra have been sorted such that the position of 
a spectrum in this figure roughly corresponds to the position of the star 
on the sky, i.e., north is up and east is to the left.
{\it (b)\/} Same as {\it (a)\/} but for stars toward h~Per (stars
marked with boxes in Fig.~1).  Stars marked with an ``h'' have been
identified as cluster members by Slesnick, Hillenbrand, \& Massey
(2002).}

\end{figure}

\clearpage

\begin{figure}
\epsscale{0.65}
\figurenum{2}
\plotone{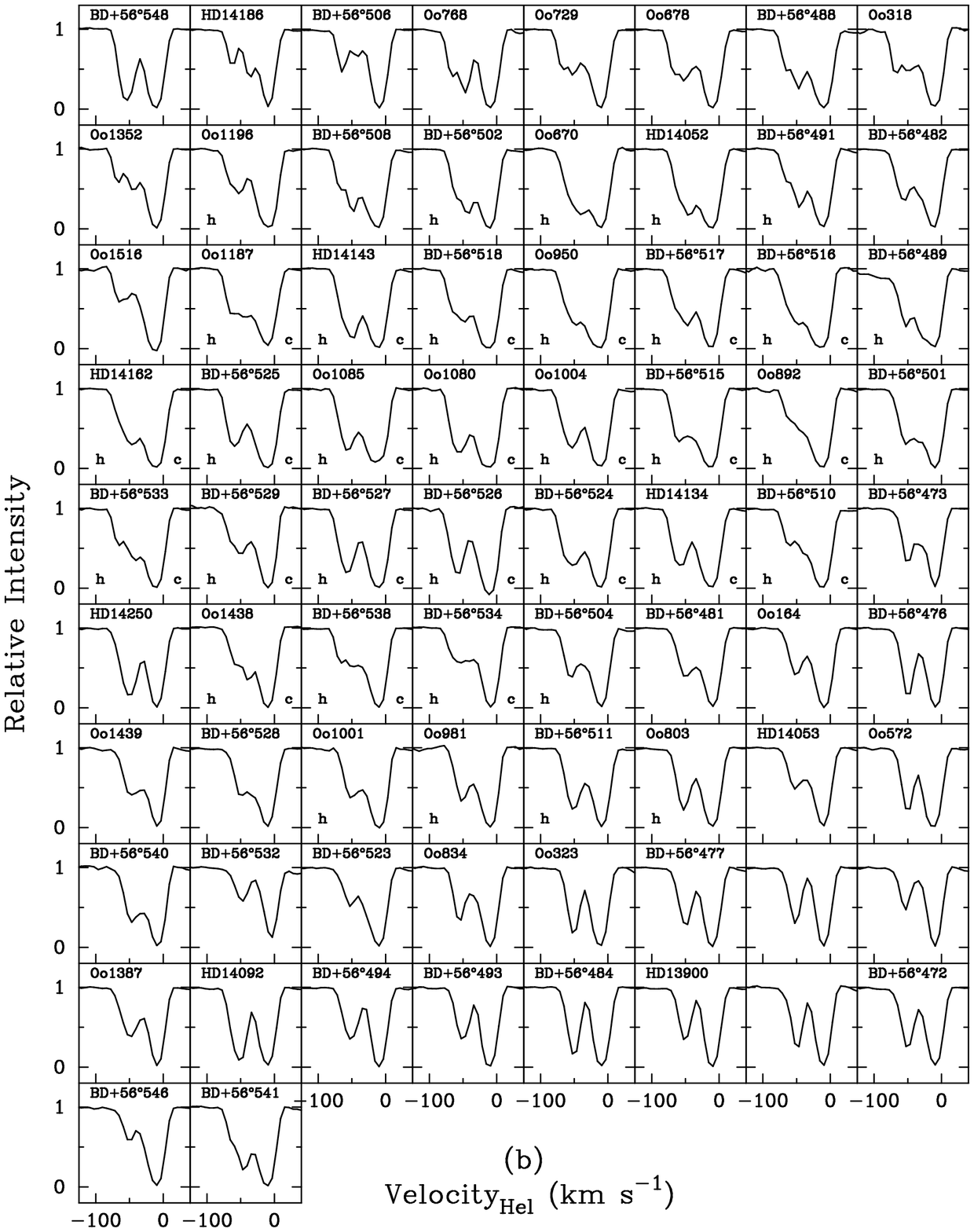}
\caption{(Cont.)}
\end{figure}

\clearpage

\begin{figure}
\epsscale{0.7}
\figurenum{3}
\plotone{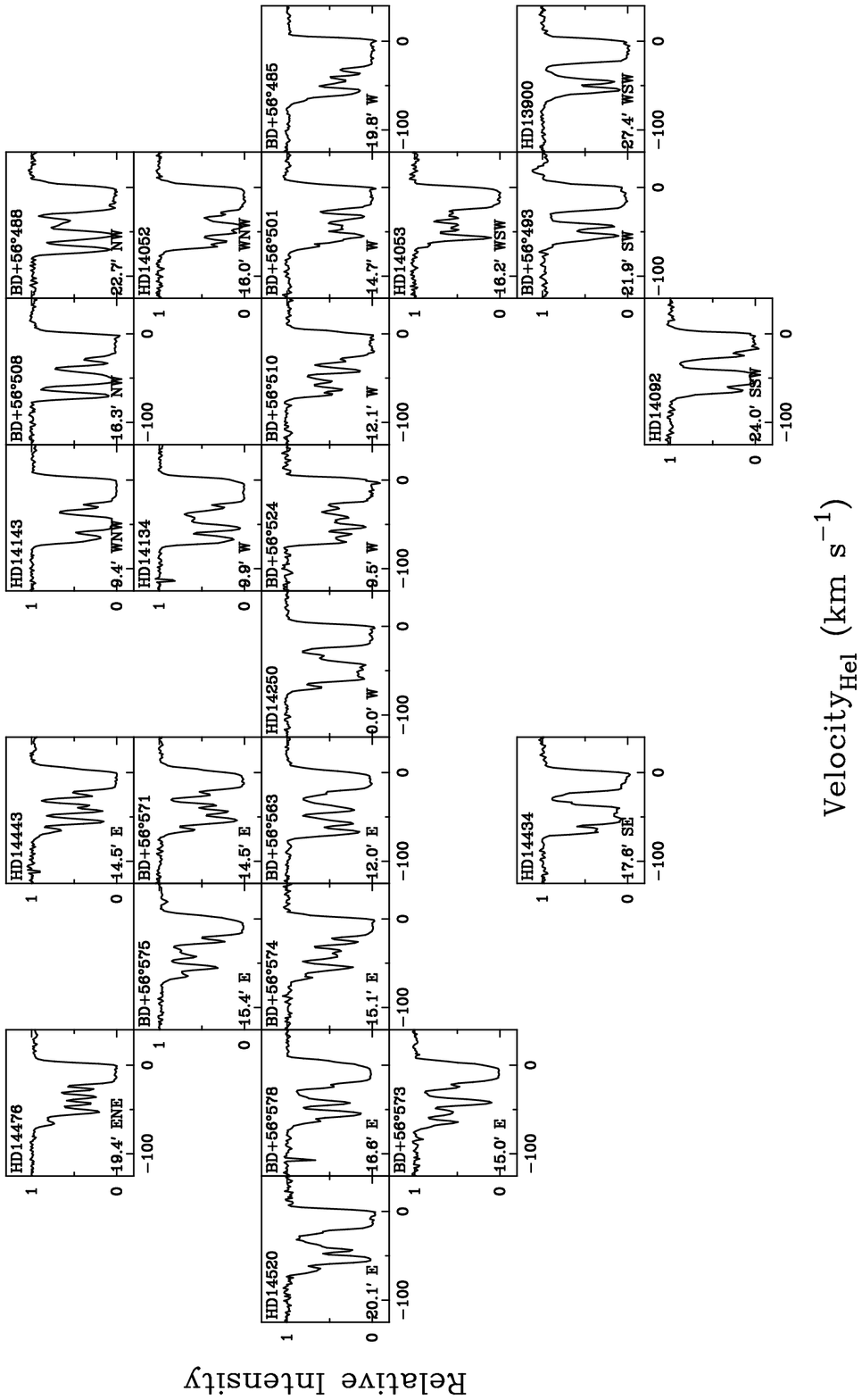}
\caption{High-resolution (3~\kms) KPNO Coud\'e Feed spectra of 
interstellar Na~{\sc i}~D$_2$ (5889\AA) absorption seen for stars 
toward h and $\chi$~Per.  These spectra have been labeled with the
identification of the background star.  The coordinate offset 
given in each panel indicates relative position of the star with 
respect to HD~14250, which lies between the cores of both clusters.}
\end{figure}

\clearpage

\begin{figure}
\epsscale{0.8}
\figurenum{4}
\plotone{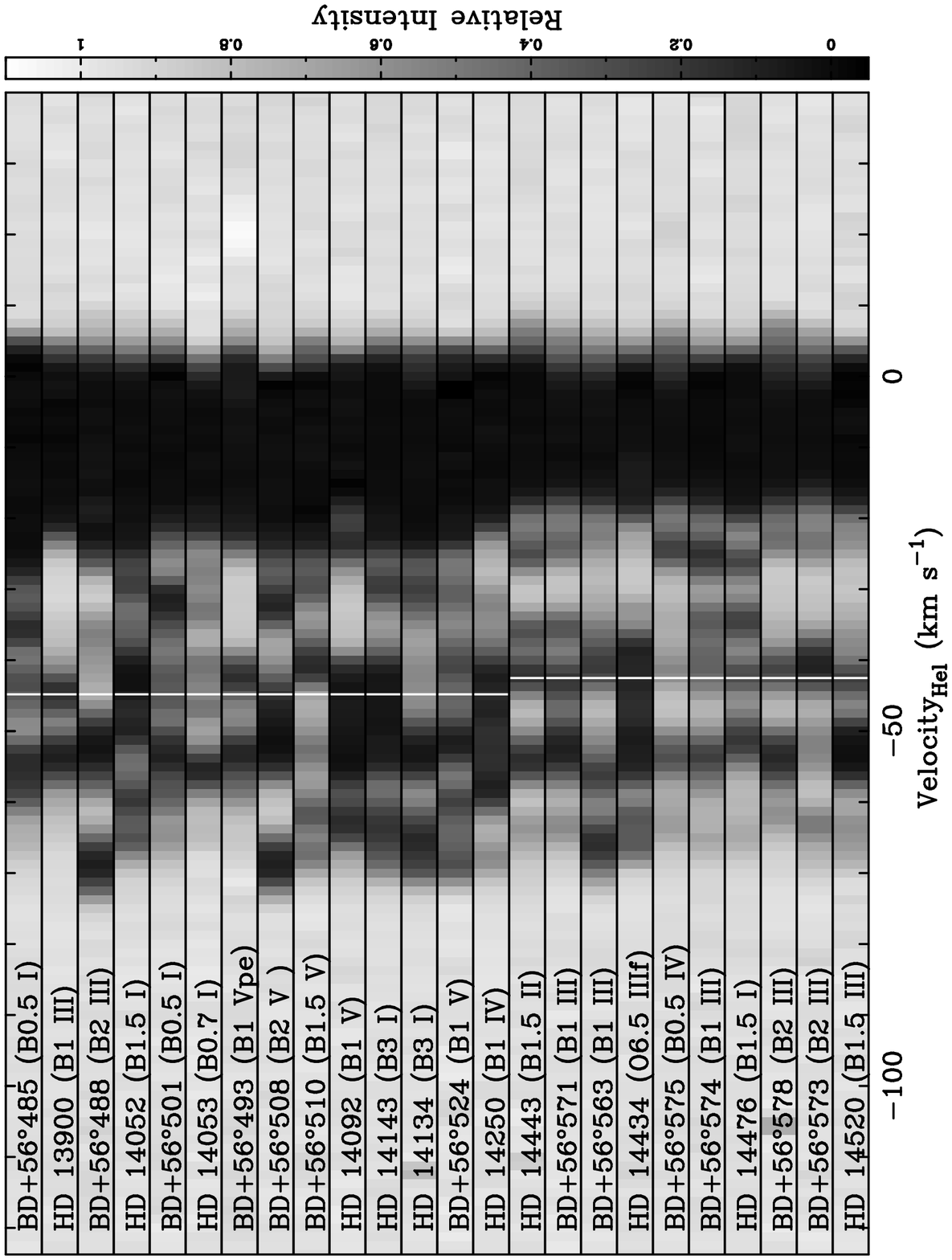}
\caption{Grey-scale images of our KPNO Coud\'e Feed spectra of interstellar
Na~{\sc i}~D$_2$ (5889\AA) absorption.  The spectral images have been 
arranged in order of increasing Right Ascension from top to bottom.  Each 
spectrum is labeled with the identification of the background star and 
its spectral type as given by Slesnick, Hillenbrand, \& Massey (2002).
The white lines in the spectral images at $-44.8$~\kms\ and $-42.5$~\kms\
show the mean heliocentric radial velocities of h and $\chi$~Per, as 
determined by Liu, Janes, \& Bania (1989), respectively,}
\end{figure}

\clearpage

\begin{figure}
\epsscale{0.6}
\figurenum{5}
\plotone{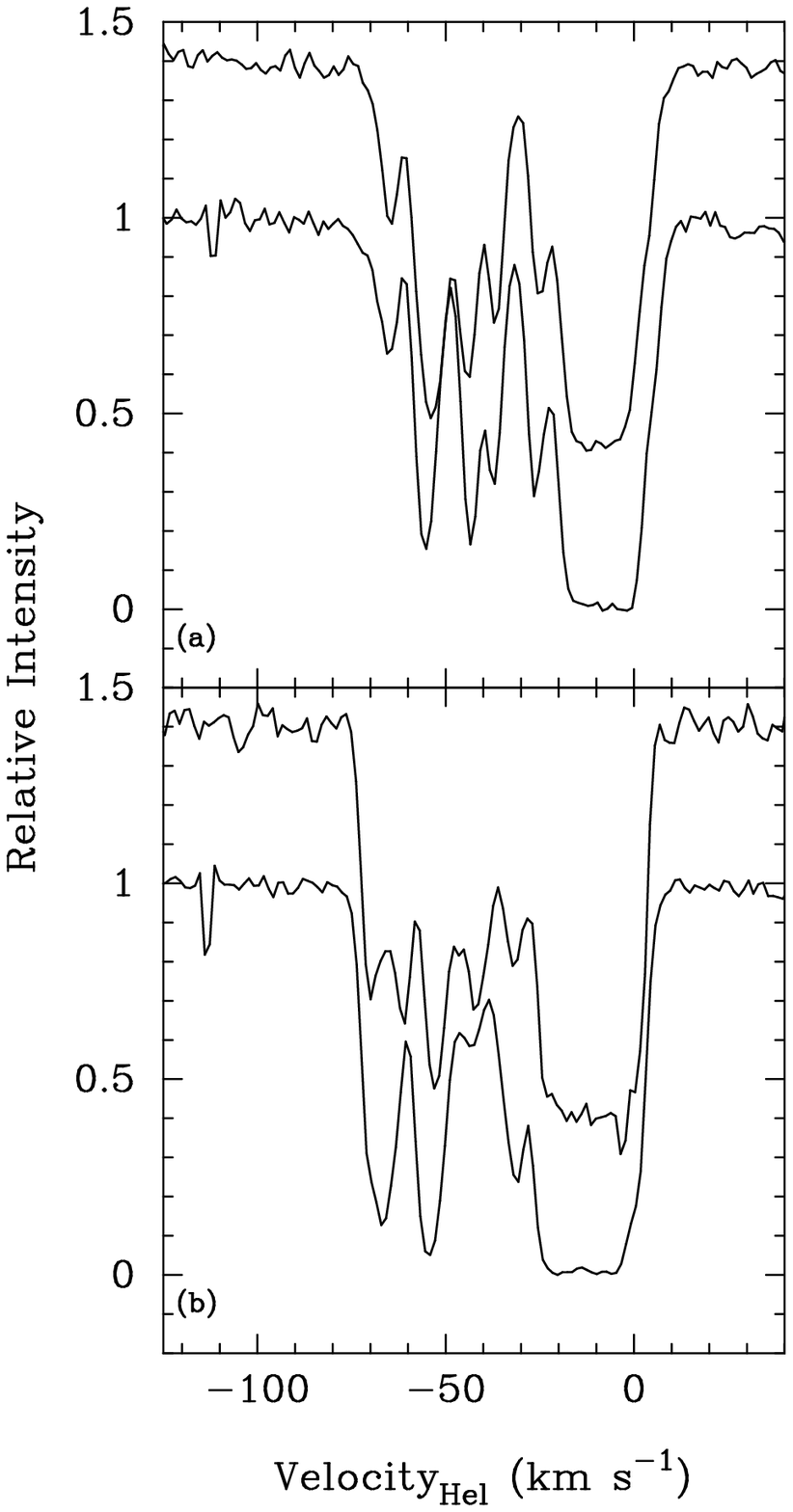}
\caption{High-resolution (3~\kms) KPNO Coud\'e Feed spectra of the
interstellar \sodii\ absorption toward {\it (a)} HD~14443 and 
BD$+$56\arcdeg 571 and {\it (b)} HD~14134 and BD$+$56\arcdeg 524 
that shows the differences and similarities between nearby sightlines.
In both {\it (a)} and {\it (b)} the spectra of BD$+$56\arcdeg 571 and 
BD$+$56\arcdeg 524 have been shifted along the y-axis by 0.4 to aid in the
comparison of spectra.}
\end{figure}

\clearpage

\begin{figure}
\epsscale{0.8}
\figurenum{6}
\plotone{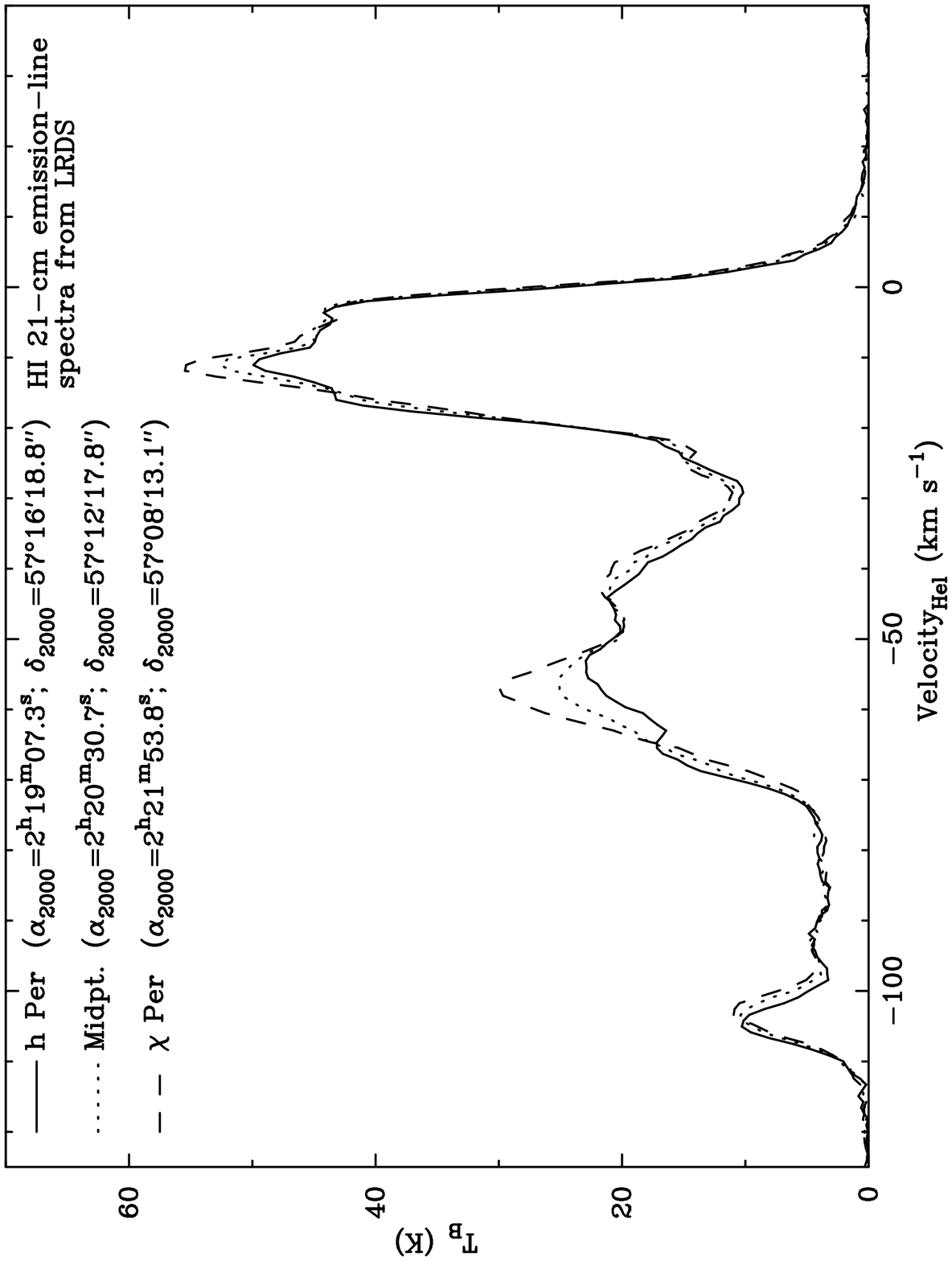}
\caption{\hi\ 21-cm emission line spectra from the Low Resolution DRAO
Survey (LRDS) of the Galactic plane (Higgs \& Tapping 2000).  The
spectra have been extracted from the survey data at positions that
roughly correspond to the core of h~Per (solid), the core of
$\chi$~Per (dashed), and a location between the clusters (dotted).}
\end{figure}

\clearpage

\begin{figure}
\epsscale{0.7}
\figurenum{7}
\plotone{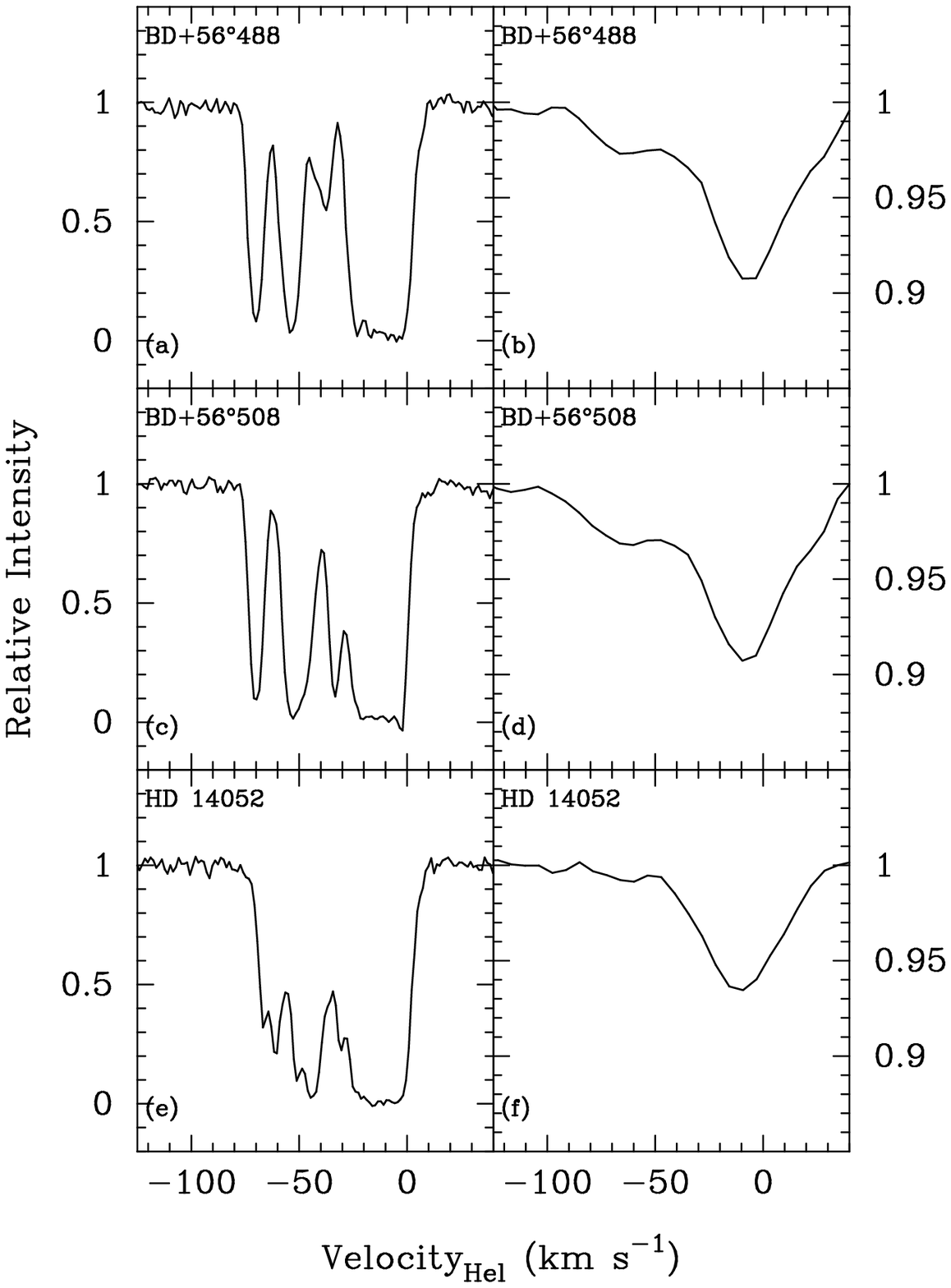}
\caption{{\it (a)} High-resolution (3~\kms) KPNO Coud\'e Feed spectrum
of interstellar \sodii\ absorption toward BD$+$56\arcdeg 488. {\it (b)}
Moderate resolution (12~\kms) WIYN Hydra spectrum of $\lambda 5797$ DIB
absorption toward BD$+$56\arcdeg 488. {\it (c)} Same as {\it (a)} toward
BD$+$56\arcdeg 508. {\it (d)} Same as {\it (b)} toward BD$+$56\arcdeg 508. 
{\it (e)} Same as {\it (a)} toward HD~14052. {\it (f)} Same as {\it (b)}
toward HD~14052.  The strong interstellar \sodii\ absorption seen in the
spectra of BD$+$56\arcdeg 488 and BD$+$56\arcdeg 508 at $V_{Hel} \sim 
-70$~\kms\ appears to be associated with a corresponding feature in the
$\lambda 5797$ DIB spectra.  Neither the strong $V_{Hel} \sim -70$~\kms\
interstellar \sodii\ absorption nor the enhanced $\lambda 5797$ DIB absorption
are detected toward other stars in our sample, such as HD~14052.}
\end{figure}


\clearpage

\begin{thebibliography}{}

\bibitem[Andrews et al. (2001)]{aml01} Andrews, S. M., Meyer, D. M., \&
Lauroesch, J. T. 2001, ApJ, 552, L73

\bibitem[Argelander (1903)]{bdcat} Argelander, F. W. A. 1903, Bonner
Durchmusterung, Vol. 2, ed. A. Marcus and E. Weber (Bonn: Verlag)

\bibitem[Bates et al. (1995)]{bates95} Bates, B., Kemp, S. N., Keenan,
F. P., \& Davies, R. D. 1995, ApJ, 444, 672

\bibitem[Bates et al. (1993)]{bkm93} Bates, B., Kemp, S. N., \& Montgomery,
A. S. 1993, A\&AS, 97, 937

\bibitem[Beals (1936)]{beals36} Beals, C. S. 1936, MNRAS, 96, 661

\bibitem[Campbell \& Albrecht (1909)]{ca09} Campbell, W. W., \& Albrecht,
S. 1909, Lick Obs. Bull., 5, 174

\bibitem[Cannon \& Pickering (1918)]{hdcat} Cannon, A. J., \& Pickering,
E. C. 1918--1924, Ann. Astron. Obs. Harvard College, 91-99

\bibitem[Cho et al. (2002)]{clv02} Cho, J., Lazarian, A., \& Vishniac,
E. T. 2002, ApJ, 566, L49

\bibitem[Crawford (2001)]{craw01} Crawford, I. A. 2001, MNRAS, 328, 1115

\bibitem[Crawford (2002)]{craw02} Crawford, I. A. 2002, MNRAS, 334, L33

\bibitem[Crawford et al. (1989)]{craw89} Crawford, I. A., Barlow, M. J., 
\& Blades, J. C. 1989, 336, 212

\bibitem[Crawford et al. (2000)]{craw00} Crawford, I. A., Howarth, I. D.,
Ryder, S. D., \& Stathakis, R. A. 2000, MNRAS, 319, L1

\bibitem[Deshpande (2000)]{d00} Deshpande, A. A. 2000, MNRAS, 317, 199

\bibitem[Diamond et al. (1989)]{diamond89} Diamond, P. J., Goss, W. M., 
Romney, J. D., Booth, R. S., Kalberla, P. M. W., \& Meabold, U. 1989, ApJ,
347, 302

\bibitem[Dickel \& Wall (1974)]{dw74} Dickel, H. R., \& Wall, J. V. 1974,
A\&A, 31, 5

\bibitem[Dieter, Welch, \& Romney (1976)]{dwr76} Dieter, N. H., Welch, W. J.,
\& Romney, J. D. 1976, ApJ, 206, L113

\bibitem[Eddington (1926)]{edd26} Eddington, A. S. 1926, Observatory, 49, 193

\bibitem[Elmegreen (1999)]{be99} Elmegreen, B. G. 1999, ApJ, 527, 266

\bibitem[ESA (1997)]{esa97} ESA 1997, The Hipparcos and Tycho Catalogues,
ESA SP-1200, ESA Publications Division, Noordwijk

\bibitem[Faison \& Goss (2001)]{fg2001} Faison, M. D., \& Goss, W. M.
2001, AJ, 121, 2706

\bibitem[Faison et al. (1998)]{faison98} Faison, M. D., Goss, W. M., 
Diamond, P. J., \& Taylor, G. B. 1998, AJ, 116, 2916

\bibitem[Foing \& Ehrenfreund (1997)]{fe97} Foing, B. H., \& Ehrenfreund, 
P. 1997, A\&A, 317, L59

\bibitem[Frail et al. (1994)]{frail94} Frail, D. A., Weisberg, J. M.,
Cordes, J. M. \& Mathers, C. 1994, ApJ, 436, 144

\bibitem[Garc\'{\i}a \& Walborn (2000)]{gw00} Garc\'{\i}a, B., \&
Walborn, N. R. 2000, PASP, 112, 1549 

\bibitem[Gardner, Dickel, \& Whiteoak (1973)]{gdw73} Gardner, F. F.,
Dickel, H. R., \& Whiteoak, J. B. 1973, A\&A, 7, 349

\bibitem[Georgelin \& Georgelin (1976)]{gg76} Georgelin, Y. P., \&
Georgelin, Y. M. 1976, A\&A, 49, 57

\bibitem[Gibson et al. (2000)]{gthd2000} Gibson, S. J., Taylor, R.,
Higgs, L. A., \& Dewdney, P. E. 2000, ApJ, 540, 851

\bibitem[Harper (1914)]{harper14} Harper, W. E. 1914, Publ. Dom. Obs. 
Ottowa, 1, 335

\bibitem[Hartmann (1904)]{hartmann04} Hartmann, J. 1904, ApJ, 19, 268

\bibitem[Heger (1919)]{heger19} Heger, M. L. 1919, Lick Obs. Bull., 10, 59

\bibitem[Heger (1922)]{heger22} Heger, M. L. 1922, Lick. Obs. Bull., 10, 141

\bibitem[Heiles (1997)]{heiles97} Heiles, C. 1997, ApJ, 481, 193

\bibitem[Higgs \& Tapping (2000)]{ht00} Higgs, L. A., \& Tapping, K. F.
2000, AJ, 120, 2471

\bibitem[Hobbs (1974)]{hobbs74} Hobbs, L. M. 1974, ApJ, 191, 395

\bibitem[Hobbs (1978)]{hobbs78} Hobbs, L. M. 1978, ApJS, 38, 129

\bibitem[Horne (1986)]{horne86} Horne, K. 1986, PASP, 98, 609

\bibitem[Humphreys (1978)]{hump78} Humphreys, R. M. 1978, ApJS, 38, 309

\bibitem[Johnston et al. (2003)]{jkww03} Johnston, S., Koribalski, B., 
Wilson, W., \& Walker, M. 2003, MNRAS, 341, 941

\bibitem[Kennedy et al. (1996)]{kbk96} Kennedy, D. C., Bates, B., \& Kemp,
S. N. 1996, A\&A, 309, 109

\bibitem[Kennedy et al. (1998)]{kbk98} Kennedy, D. C., Bates, B., \& Kemp,
S. N. 1998, A\&A, 336, 315

\bibitem[Lauroesch \& Meyer (1999)]{lm99} Lauroesch, J. T., \& Meyer,
D. M. 1999, ApJ, 519, L181

\bibitem[Lauroesch \& Meyer (2003)]{lm03} Lauroesch, J. T., \& Meyer,
D. M. 2003, ApJ, 591, L123

\bibitem[Lauroesch et al. (2000)]{lmb2000} Lauroesch, J. T., Meyer, D. M.,
\& Blades, J. C. 2000, ApJ, 543, L43

\bibitem[Lauroesch et al. (1998)]{lmwb98} Lauroesch, J. T., Meyer, D. M.,
Watson, J. K., \& Blades, J. C. 1998, ApJ, 507, L89

\bibitem[Lindblad (1927)]{lindblad27} Lindblad, B. 1927, MNRAS, 87, 553

\bibitem[Liu et al. (1989)]{liu89} Liu, T., Janes, K. A., \& Bania,
T. M. 1989, AJ, 98, 626

\bibitem[Marsh (1989)]{marsh89} Marsh, T. 1989, PASP, 100, 1032

\bibitem[McCarthy et al. (2000)]{mctt00} McCarthy, M. C., Chen, W., 
Travers, M. J. \& Thaddeus, P. 2000, ApJS, 113, 105

\bibitem[Meyer \& Blades (1996)]{mb96} Meyer, D. M., \& Blades, J. C.
1996, 464, L179

\bibitem[Meyer \& Lauroesch (1999)]{ml99} Meyer, D. M., \& Lauroesch, J. T.
1999, ApJ, 464, L179 

\bibitem[Morgan, Sharpless, \& Osterbrock (1952)]{mso52} Morgan, W. W.,
Sharpless, S., \& Osterbrock, D. 1952, AJ, 57, 3

\bibitem[M\"unch (1953)]{munch53} M\"unch, G. 1953, \pasp, 65, 179

\bibitem[M\"unch (1957)]{munch57} M\"unch, G. 1957, ApJ, 125, 42

\bibitem[O'Dell et al. (1993)]{odell93} O'Dell, C. R., Valk, J. H., Wen,
Z., \& Meyer, D. M. 1993, ApJ, 403, 678

\bibitem[Oosterhoff (1937)]{oo37} Oosterhoff, P. Th. 1937, Ann. Sternw. 
Leiden, 17, 1

\bibitem[Pan et al. (2001)]{pfw2001} Pan, K., Federman, S. R., \& Welty,
D. E. 2001, ApJ, 558, L105

\bibitem[P\'equignot \& Aldrovandi (1986)]{pa86} P\'equignot, D., \& 
Aldrovandi, S. M. V. 1986, A\&A, 161, 169

\bibitem[Plaskett (1923)]{plaskett23} Plaskett, J. S. 1923, MNRAS, 84, 80

\bibitem[Plaskett \& Pearce (1930)]{pp30} Plaskett, J. S., \& Pearce, J. A.
1930, MNRAS, 90. 243

\bibitem[Price et al. (2000)]{pcb2000} Price, R. J., Crawford, I. A., \&
Barlow, M. J. 2000, MNRAS, 312, L43

\bibitem[Price et al. (2001b)]{pcbh2001b} Price, R. J., Crawford, I. A.,
Barlow, M. J., \& Howarth, I. D. 2001b, MNRAS, 328, 555

\bibitem[Price et al. (2001a)]{pch2001} Price, R. J., Crawford, I. A., \&
Howarth, I. D. 2001a, MNRAS, 321, 553

\bibitem[Rollinde et al. (2003)]{rbfp2003} Rollinde, E., Boiss\'e, P.,
Federman, S. R., \& Pan, K. 2003, A\&A, 401, 215

\bibitem[Salama et al. (1996)]{sbat96} Salama, F., Bakes, E. L. O., 
Allamandola, L. J., \& Tielens, A. G. G. M 1996, ApJ, 458, 621

\bibitem[Slesnick et al. (2002)]{shm02} Slesnick, C. L., Hillenbrand,
L. A., \& Massey, P. 2002, ApJ, 576, 880

\bibitem[Smoker et al. (2001)]{smok2001} Smoker, J. V., Lehner, N., 
Keenan, F. P., Totten, E. J., Murphy, E., Sembach, K. R., Davies, R. D., 
\& Bates, B. 2001, MNRAS, 322, 13

\bibitem[Spitzer (1978)]{ppism} Spitzer, L. 1978, Physical Processes in 
the Interstellar Medium (New York: Wiley)

\bibitem[Stanimirovi\'c et al. 2003]{ss03} Stanimirovi\'c, S., Weisberg,
J. M., Hedden, A., Devine, K., Greent, T., \& Anderson, S. B. 2003,
astro-ph/0307002 

\bibitem[Struve (1928)]{struve28} Struve, O. 1928, ApJ, 67, 353

\bibitem[van Leeuwen (1999)]{vl99} van Leeuwen, F. 1999, A\&A, 341, L71

\bibitem[Walborn (1982)]{wal82} Walborn, N. R. 1982, ApJS, 48, 145

\bibitem[Walborn (1995)]{wal95} Walborn, N. R. 1995, Rev. Mexicana
Astron. Astrof. Ser. de Conf., 2, 51

\bibitem[Walborn \& Hesser (1975)]{wh75} Walborn, N. R., \& Hesser,
J. E. 1975, ApJ, 199, 535

\bibitem[Watson \& Meyer (1996)]{wm96} Watson, J. K., \& Meyer, D. M.
1996, ApJ, 474, L127

\bibitem[Welty \& Fitzpatrick (2001)]{wf01} Welty, D. E., \& Fitzpatrick,
E. L. 2001, ApJ, 551, L175

\bibitem[Welty et al. (1994)]{whk94} Welty, D. E., Hobbs, L. M., \&
Kulkarni, V. P. 1994, ApJ, 436, 152

\bibitem[Welty et al. (1996)]{welty96} Welty, D. E., Morton, D. C., \& 
Hobbs, L. M. 1996, ApJS, 106, 533

\bibitem[White et al. (2001)]{w01} White, R. E., Allen, C. L., 
Forrester, W. B., Gonnella, A. M., \& Young, K. L. 2001, ApJS,
132, 253

\bibitem[Whiteoak \& Gardner (1980)]{wg80} Whiteoak, J. B., \& Gardner, 
F. F. 1980, PASA, 4, 95

\bibitem[Whiteoak \& Gardner (1985)]{wg85} Whiteoak, J. B., \& Gardner, 
F. F. 1985, PASA, 6, 164

\bibitem[Wilson \& Merrill (1937)]{wm37} Wilson, O. C., \& Merrill, P. W.
1937, ApJ, 86, 44

\bibitem[Young (1920)]{young20} Young, R. K. 1920, Pub. Dom. Astro. Obs.,
1, 219

\end{thebibliography}
\end{document}